\documentclass[particles,article,accept,moreauthors,pdftex]{Definitions/mdpi} 
\firstpage{1} 
\pubvolume{xx}
\issuenum{1}
\articlenumber{5}
\pubyear{2020}
\copyrightyear{2020}
\history{Received: 28 February 2020; Accepted: 24 March 2020; Published: date}
\updates{yes} 

\usepackage{upgreek}
\usepackage{textcomp}

\Title{A Study of the Properties of the QCD Phase Diagram in High-Energy Nuclear Collisions}

\newcommand{\sNN}{$\sqrt{s_{NN}}$}
\newcommand{\KV}{{\mbox{$\kappa \sigma^{2}$}}}
\newcommand{\SD}{{\mbox{$S \sigma$}}}
\newcommand{\VM}{{\mbox{$\sigma^{2}/M$}}}

\Author{Xiaofeng Luo $^{1}$, Shusu Shi $^{1,}$*$^{,\dagger}$, Nu Xu $^{1,2}$ and Yifei Zhang $^{3}$}
\AuthorNames{Firstname Lastname, Firstname Lastname and Firstname Lastname}

\address{%
$^{1}$ \quad Key Laboratory of Quark \& Lepton Physics (MOE) and Institute of Particle Physics, Central China Normal University, Wuhan 430079, China; xfluo@mail.ccnu.edu.cn (X.L.); nxu@lbl.gov(N.X.)\\
$^{2}$ \quad Institute of Modern Physics, Chinese Academy of Sciences, Lanzhou 730000, China\\
$^{3}$ \quad State Key Laboratory of Particle Detection and Electronics, University of Science and Technology of China, Hefei 230026, China; ephy@ustc.edu.cn}
\corres{Correspondence: shiss@mail.ccnu.edu.cn}
\firstnote{The authors contribute equally to this paper.}

\abstract{With the aim of understanding the phase structure of nuclear matter created in high-energy nuclear collisions at finite baryon density, a beam energy scan program has been carried out at Relativistic Heavy Ion Collider (RHIC).
 In~this mini-review, most recent experimental results on collectivity, criticality and heavy flavor productions will be discussed. The~goal here is to establish the connection between current available data and future heavy-ion collision experiments in a high baryon density region. }


\begin{document}

\section{Introduction}
 
Most of the visible matter in our universe can be described by the Quantum Chromdynamics (QCD), the~standard theory of strong interactions. 
In the beginning of the century, the~new form of matter, the~quark-gluon plasma (QGP) in which quarks and gluons are `freed' 
in a much larger volume compared to that of nucleon's, was discovered in the largest heavy-ion colliders RHIC and 
LHC
~\cite{BRAHMS:qgp, PHOBOS:qgp, STAR:qgp, PHENIX:qgp} at vanishing baryonic density. 
Soon after the discovery, a~serious question was asked: what is the structure of the nuclear matter at high baryonic density? 

\begin{figure}[H]
\centering
\includegraphics[width=10 cm]{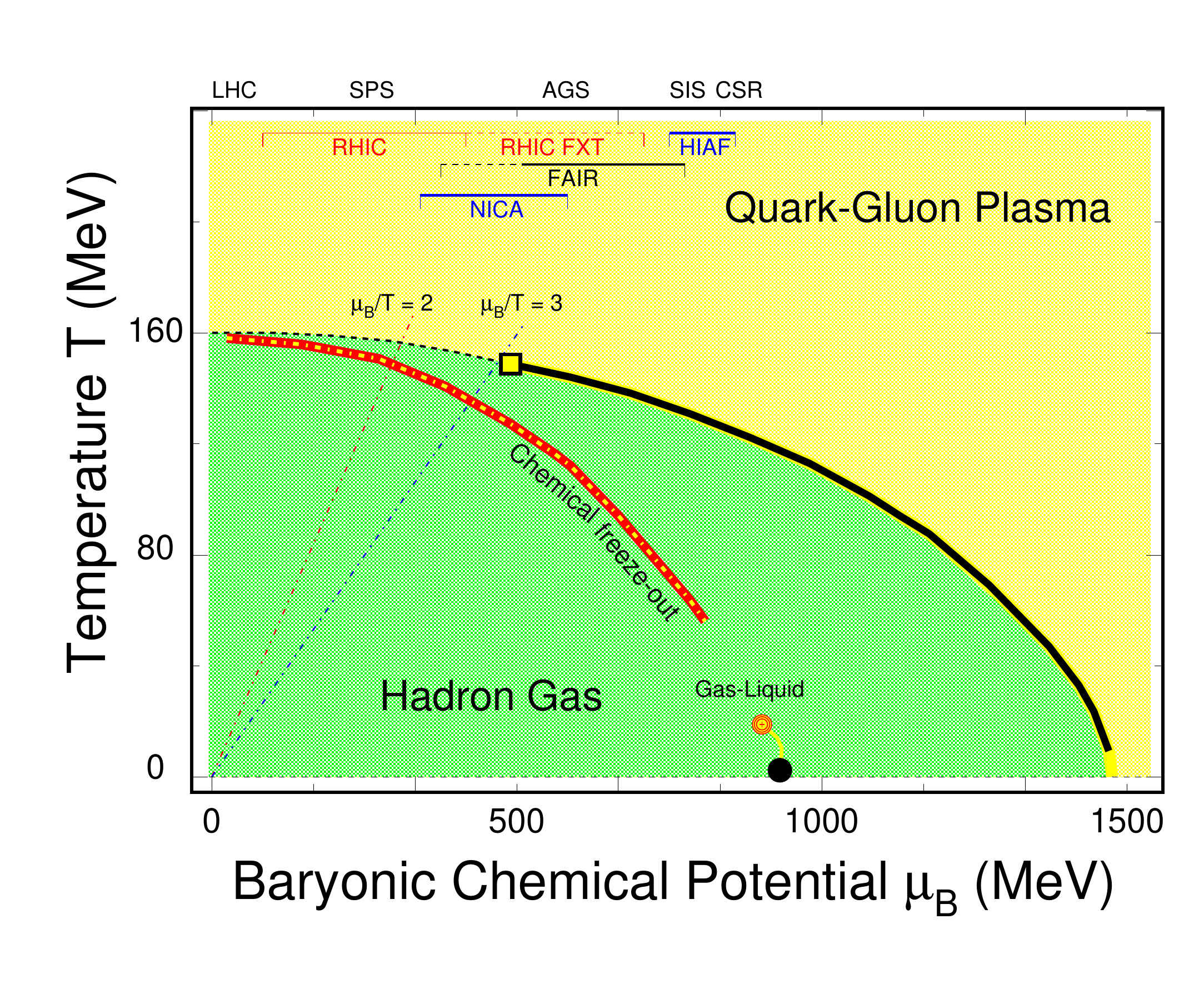}
\caption{(Color online) Schematic Quantum Chromdynamics (QCD) phase diagram in the thermodynamic parameter space spanned by the temperature $T$ and baryonic-chemical potential $\mu_B$. 
Experimentally extracted chemical freeze-out parameters are shown as the red-line~\cite{beta2}. 
The~dashed-line near $\mu_B =$ 0 indicates the crossover transition from the Quark-Gluon-Plasma to Hadron Gas~\cite{Aoki,Bazavov:2011nk,Bazavov:2018mes,Bellwied:2015rza,Borsanyi:2020fev}. 
The solid-line and the square show the expected 1st-order phase boundary and its end point: critical point. 
The regions of $\mu_B/T < 2$ and $< 3$ are shown as red and blue dot-dashed lines, respectively. 
The Lattice QCD calculations have concluded that the QCD critical point is unlikely to exist at $\mu_B/T < 2.5$~\cite{Bazavov1, Bazavov2}. 
The red-circle and the yellow-line represent the liquid-gas transition~\cite{QCDreview}. 
Regions for physics programs of RHIC beam energy scan and fixed-target (US)~\cite{stardrupal}, NICA (Russia)~\cite{nica}, FAIR (Germany)~\cite{cbm} as well as the HIAF (China)~\cite{hiaf} are indicated at the top of the plot.}\label{fig1_tmu_dl_dec2019}
\end{figure}

Tremendous efforts from both experimental and theoretical sides have launched in order to address the question. 
Figure~\ref{fig1_tmu_dl_dec2019} summarizes the current status of the studies. At~the zero baryonic density, the~transition from QGP to hadronic matter 
is a smooth-crossover at $T$ = $150-160$ MeV~\cite{Aoki,Bazavov:2011nk,Bazavov:2018mes,Bellwied:2015rza,Borsanyi:2020fev}, see dashed-line in the figure. 
These results are extracted from the state of the art Lattice gauge theory calculations. 
At high baryonic density, on~the other hand, one would expect a first-order phase transition, shown as a black solid-line. 
Thermodynamically the first-order phase boundary line must end at finite baryonic density, 
this is the illusive QCD critical point (CP). Again, recent Lattice calculations have concluded that the QCD critical is `unfavored'~\cite{Bazavov1, Bazavov2} when $\mu_B/T < 2.5$. The~red-line is the chemical freeze-out curve extracted from the measured 
hadron yields. The~collision energies with the corresponding accelerator complex are indicated 
at the top of the~figure. 

\begin{figure}[H]
\centering
\includegraphics[width=14 cm]{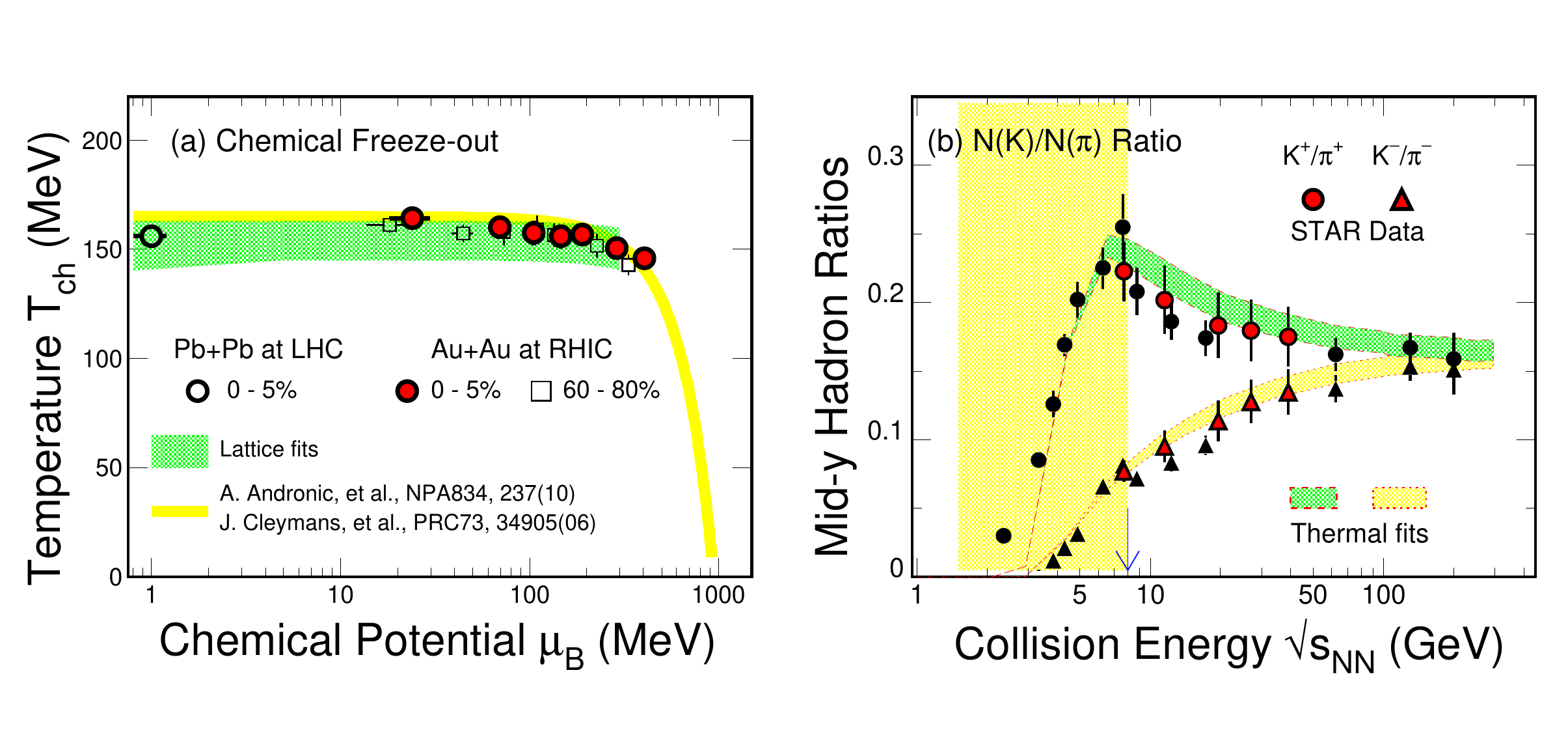}
\caption{(Color online) (\textbf{a}) Experimental results of chemical freeze-out temperature as a function of the baryonic chemical potential from the RHIC BES-I~\cite{beta2} and LHC~\cite{alicedata1, alicedata2}. Red-circles and black-squares represent results from the top $5\%$ and 60--80\% Au+Au collisions at RHIC, respectively. Open-circle is from the 2.76 TeV 0--5\% Pb+Pb collisions at LHC. The~hatched green-band represents the Lattice results~\cite{Kaczmarek:2011zz,Bazavov:2018mes}. The~yellow-line shows the empirical thermal fits results~\cite{Andronic, Cleymans}. (\textbf{b}) Mid-rapidity particles of positive and negative kaons over pions shown as circles and triangles, respectively. The~results~\cite{Chatterjee} of thermal model fits are shown as hatched-bands in the plot and the high-baryon-density region is highlighted in yellow. 
}\label{fig2_plot2_cfo_kpi_dec2019}
\end{figure}

Experimental status on the beam energy scan (BES) is highlighted in Figure~\ref{fig2_plot2_cfo_kpi_dec2019}. Plot (a) shows the chemical freeze-out temperature 
$T_{\rm ch}$ as a function of the baryonic chemical potential $\mu_{B}$. Both ALICE at LHC and STAR at RHIC 
results 
clearly show that at the vanishing baryon density, i.e.,~at high collision energy, the~data driven freeze-out temperature is consistent 
with the Lattice calculation, $T_{ch} \sim 160$ MeV. Many authors have tried to analyze the chemical freeze-out conditions. Those include the results fluctuations analysis from Lattice QCD~\cite{Bazavov:2012vg,Borsanyi:2014ewa,Bazavov:2015zja,Alba:2014eba,Bluhm:2018aei}, HRG model~\cite{beta2} and other methods~\cite{Poberezhnyuk:2019pxs,Alba:2020jir}. In~the low baryon density region, $\mu_{B} \le 400$ MeV, the~$\mu_{B}$ dependence of the freeze-out temperature is quite weak and the value of the freeze-out temperature is around 150--160 MeV. More dramatic drop of the temperature is seen in the high baryon density region. Plot (b) is the kaon over pion yield ratios, extracted from central heavy-ion collisions, as~a function of the collision energies. 
While one observes the smooth increase of the negative kaon over pion ratio with the collision energy, 
the positive ratio shows a broad peak around $\sqrt{s_{NN}}\sim 8$ GeV and eventually 
merged with the negative ratios at high collision energy, $\sqrt{s_{NN}}\ge 100$ GeV, 
where the pair production becomes dominant. Due to the associate channel, 
$N+N \rightarrow N+\Lambda+K^+$, positive kaons carry information on baryon density. 
The~peak in plot (b) implies the maximum freeze-out density reached at 8 GeV. 
Later in the discussions, we attribute the region of $2 \le$ $\sqrt{s_{NN}} \le 8$ GeV as the high baryon density region (HBDR) 
as indicated by the yellow-area in the~plot.

\unskip

Collective flow (collectivity) and the critical behavior (criticality) are important aspects in high-energy nuclear collisions. 
In this short review, we will discuss the experimental status including the results on collectivity and criticality 
from high-energy nuclear collisions. RHIC has provided most recent data so we will focus on the information. 
At the end we will address the importance of the future fixed-target experiments such as STAR fixed-target program in BES-II, 
CBM at FAIR~\cite{cbm} as well the CEE at HIAF~\cite{hiaf}. 


\section{Beam Energy Dependence of the Collectivity}

\begin{figure}[H]
\centering
\includegraphics[width=9 cm]{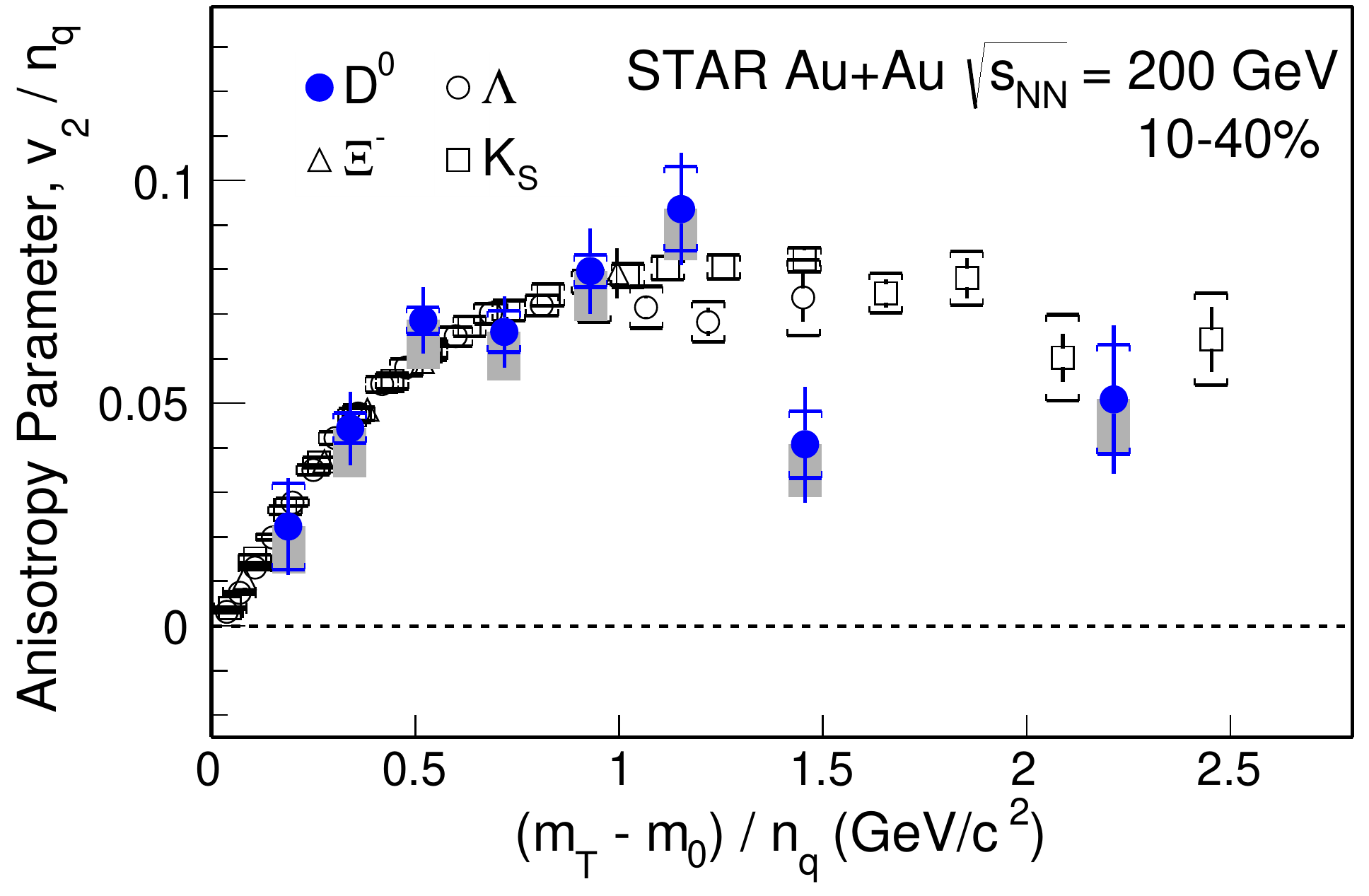}
\caption{(Color online) The number of constituent quark ($n_q$) scaled $v_2$ as a function of $(m_T - m_0)/n_q$ in 10\%--40\%  
Au+Au collisions for $D^0$, $\Xi^-$, $\Lambda$ and $K_S^0$ (from~\cite{D0v2PRL}). 
Where $m_T$ is square root of the rest mass squared plus transverse momentum squared.}\label{D0v2}
\end{figure}

The main goal of high energy heavy-ion collisions, such as collisions at LHC and top energy collisions at RHIC, is to study the properties of 
new form of matter QGP. QGP is a thermalized (or~nearly) system with partonic degree of freedom. 
The elliptic flow measurement of multi-strange hadrons and $\phi$ mesons indicates the partonic collectivity has been built up at the
top energy heavy-ion collisions at RHIC~\cite{STARmultis1,STARmultis2,STARmultis3,STARmultis4,STARmultis5,STARmultis6, STARmultis7}. The~Heavy Flavor Tracker, a~high resolution silicon detector system, 
which was installed in the year of 2013, provides high vertex position resolution. The~significance of charmed hadron reconstruction is significantly improved.
Thus the precise measurement of $D^0$ becomes possible.  Figure~\ref{D0v2} shows the $v_2$ results for $D^0$, $\Xi^-$, $\Lambda$ and $K_S^{0}$~\cite{D0v2PRL}.
A number of constituent quark ($n_q$) is tested by scaling both $v_2$ and $m_T - m_0$ with $n_q$.  A~simple quark coalescence or
recombination model suggests the baryon $v_2$ would be 1.5 times of the meson $v_2$ assuming the collectivity has been attained in
the partonic stage, as~the number of constituent quarks for baryons is 3 where it is 2 for mesons.  When discussing the $n_q$ scaling, 
we usually focus on the intermediate $p_T$ range in which the $v_2$ value saturates.
The $D^0$ $v_2$ follows the $n_q$ scalings with selected multi-strange and strange hadrons in Figure~\ref{D0v2}. 
It~indicates the collectivity of parton level has been built up from light flavor $u$, $d$ quarks to strange and charm quarks.
Since the mass of the charm-quark is much larger than the temperature reached in the system, the~observed strong charm-quark collectivity can be interpreted as the thermalization of the medium created in the 200 GeV Au+Au collisions at RHIC~\cite{D0v2PRL}. To~some extent, this result justified the phase diagram sketched in Figure~\ref{fig1_tmu_dl_dec2019}.

The main motivation of BES program is to explore the QCD phase boundary and critical point. 
The logic is straightforward:
as the collision energy decreases, the~conditions of QGP formation are no longer satisfied at some point. It offers us a unique experimental way
to investigate the QCD phase structure. 
The transverse radial flow velocity $\beta$ is obtained by fitting the transverse momentum $p_T$ spectra with a
blast wave model~\cite{blast_wave}:
\begin{equation}
\frac{dN}{p_Tdp_T} \propto \int_{0}^{R}~r~dr~m_{T}~I_{0}(\frac{p_{T}\sinh\rho(r)}{T_{\rm kin}}) \times K_{1}(\frac{m_{T}\cosh \rho(r)}{T_{\rm kin}})
\end{equation}
where $I_0$ and $K_1$ are the modified Bessel functions and $\rho(r) = \tanh^{-1} \beta$.
The model assumes a radially boosted thermalized source with two key parameters,
kinetic freeze-out temperature $T_{\rm kin}$ and a transverse collective flow velocity $\beta$.

\begin{figure}[H]
\begin{minipage}[t]{0.49\linewidth}
\centering
\includegraphics[width=0.9\textwidth]{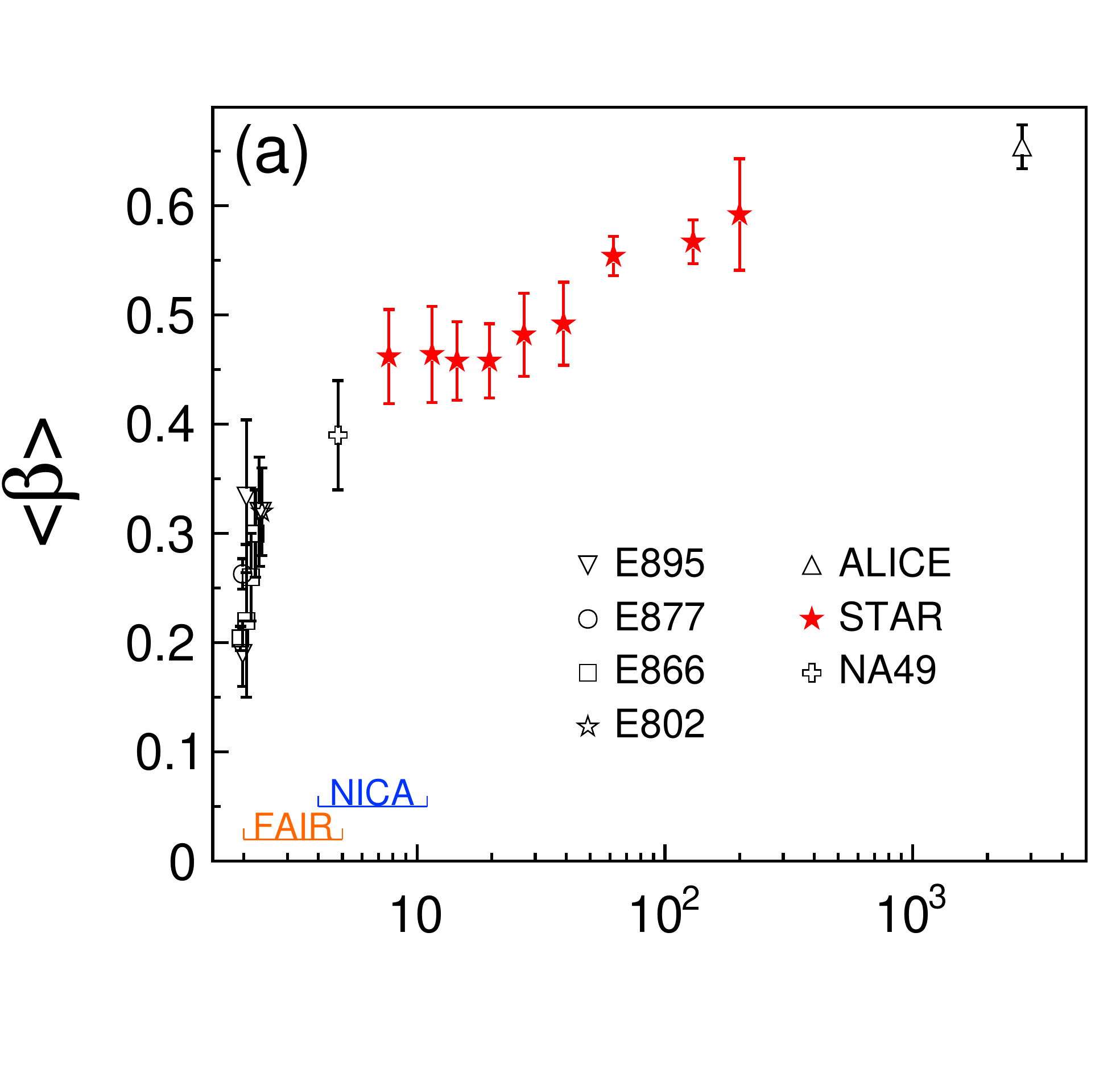}
\label{fig:side:a}
\end{minipage}
\hfill
\begin{minipage}[t]{0.49\linewidth}
\centering
\includegraphics[width=0.9\textwidth]{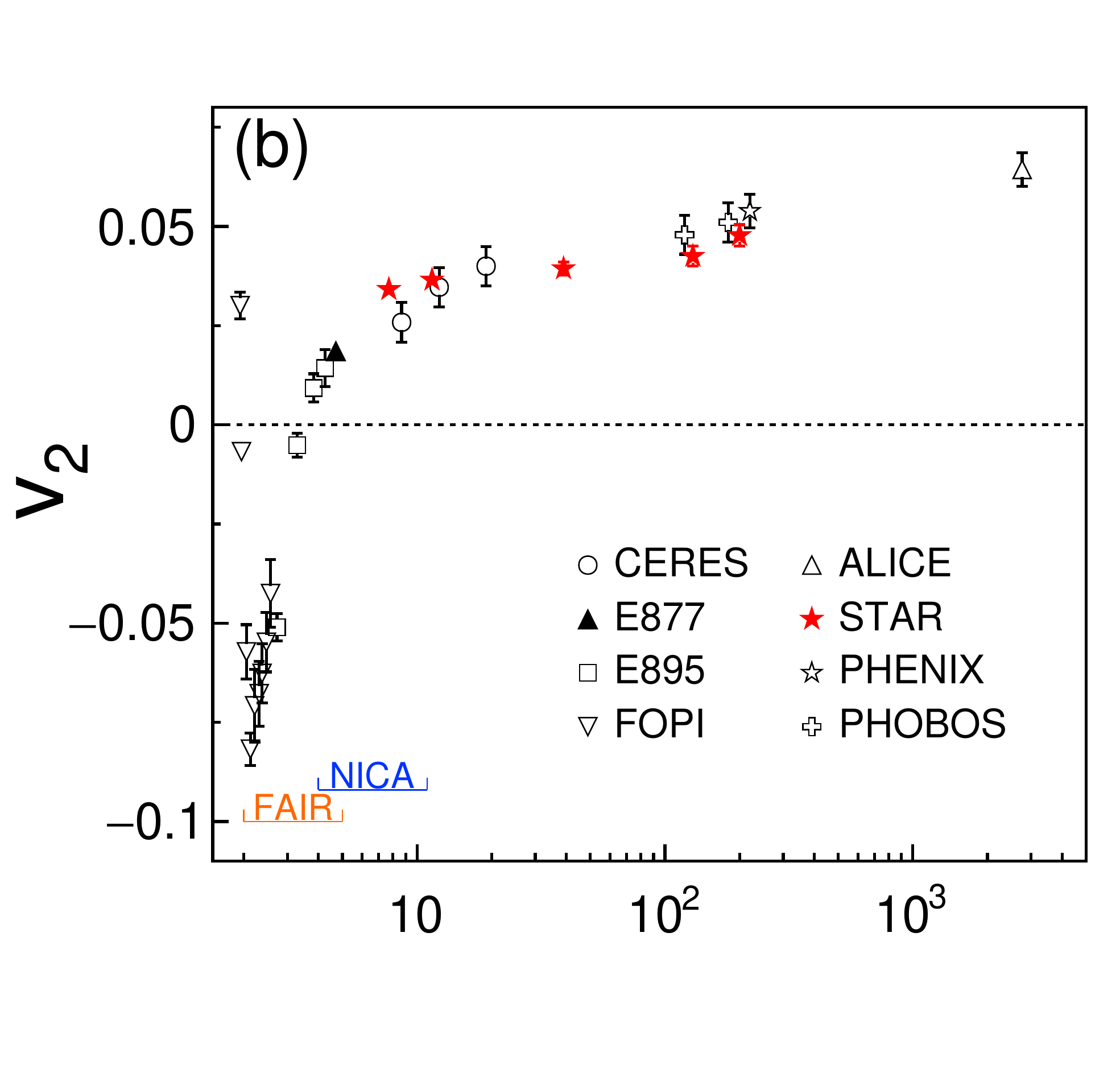}
\end{minipage}
\put(-320,10){\Large  \text { Collision Energy $\sqrt{\rm s_{NN}}$ (GeV)}}
\caption{(Color online) (\textbf{a}) The transverse radial flow velocity as a function of collision energy from 
}
\label{v0v2}
\end{figure}

Plot (a) of Figure~\ref{v0v2} shows the extracted parameter $\beta$ as a function of collision energy. 
The data points are taken from E802~\cite{beta4, beta5, beta6, beta7}, E866~\cite{beta8, beta9}, E877~\cite{beta10}, E895~\cite{beta11}, NA49~\cite{beta12, beta13, beta14, beta15}, 
STAR~\cite{beta2, beta16, beta17} and ALICE experiments~\cite{beta18} and references therein.
The $p_T$ spectra of $\pi^\pm$, $K^\pm$, $p$ and $\bar{p}$ are fitted simultaneously with the blast wave model. 
The $p_T$ range for simultaneous fitting are similar across all RHIC BES and LHC energies. 
A rapid increase of $\left \langle\beta\right \rangle$ is observed at low energies (<5 GeV),
then a steady increase follows up to LHC energy. The~six points from RHIC BES (7.7--39 GeV) are almost flat within~uncertainties.

{\captionof*{figure}{central heavy-ion 
collisions. The~data points are taken from E802~\cite{beta4, beta5, beta6, beta7}, E866~\cite{beta8, beta9}, E877~\cite{beta10}, E895~\cite{beta11}, NA49~\cite{beta12, beta13, beta14, beta15}, 
STAR~\cite{beta2, beta16, beta17} and ALICE experiments~\cite{beta18} and references therein.
The data points of RHIC and LHC are from 0--5\% central collisions. AGS and SPS energies are mostly from
0--5\% and 0--7\% central collisions respectively. 
(\textbf{b}) The $p_T$-integrated $v_2$ in 20\%--30\% most central collisions (or similar centrality) from various collision energies.
The data points are from E895~\cite{E895} for protons, NA49~\cite{NA49} for pions and FOPI~\cite{FOPI}, E877~\cite{E877}, CERES~\cite{CERES}, STAR and ALICE~\cite{alicev2} for charged hadrons. 
The RHIC results of inclusive charged particles for 130 and 200 GeV are from refs.~\cite{flowreview, runII200gevV2, STARcum, PHENIX, PHOBOS}. 
The RHIC BES data are from refs.~\cite{BESv2_1, BESv2_2}. The~energy regions  of future collider and fix-target experiments, NICA and FAIR, are indicated in both plots.}}
\vspace{12pt}

The $\left \langle\beta\right \rangle$ parameter extracted from blast wave model reflects the transverse radial flow built-up in the collision system.
The second order coefficient of final azimuth distribution in the momentum space, $v_2$, is sensitive to the initial geometry
and interactions of early stage of the collisions. It suggests that a non-monotonic variation could be observed around the
so-called, ``softest point of EOS''~\cite{v2_softest1, v2_softest2}.  The~``softest point of EOS'' is usually defined as a strong drop of
speed of sound (a minimum value) or a reduction in the pressure of the system during the dynamic evolution.
Plot (b) of Figure~\ref{v0v2} shows the $p_T$-integrated $v_2$ from 20\%--30\% or similar centrality as a function of collision energy.
The data points are from E895~\cite{E895} for protons, NA49~\cite{NA49} for pions and FOPI~\cite{FOPI}, E877~\cite{E877}, CERES~\cite{CERES}, STAR and ALICE~\cite{alicev2} for charged hadrons. The~RHIC results of inclusive charged particles for 130 and 200 GeV are from 
refs.~\cite{flowreview, runII200gevV2, STARcum, PHENIX, PHOBOS}.
The RHIC BES data are from refs.~\cite{BESv2_1, BESv2_2}.
The negative $v_2$ ($\sqrt{s_{NN}} < $3 GeV) is known due to the ``squeeze-out'' effect~\cite{FOPI}.
An increasing trend is observed for $p_T$ integrated $v_2$ from AGS to LHC.
It appears that the slope of $v_2$ with collision energy is steeper for 3--7.7 GeV 
compared to 7.7--2760 GeV,
which is consistent with that we observe for $\left \langle\beta\right \rangle$ parameter. 
The $v_2$ of charged hadrons as a function of $p_T$ does not change significantly at RHIC BES and LHC energies.
Due to the rise in mean $p_T$ which is expected from larger radial flow, the~$p_T$-integrated $v_2$ increases. 
It is consistent with the collision energy dependence of $\left \langle\beta\right \rangle$ parameter, as~discussed in panel (a) of Figure~\ref{v0v2}. 
Non-monotonic behavior which is predicted by the softening of the equation of state for a system close to the critical temperature~\cite{v2_softest1} is not~observed.

\begin{figure}[H]
\begin{minipage}[t]{0.5\linewidth}
\centering
\includegraphics[width=0.95\textwidth]{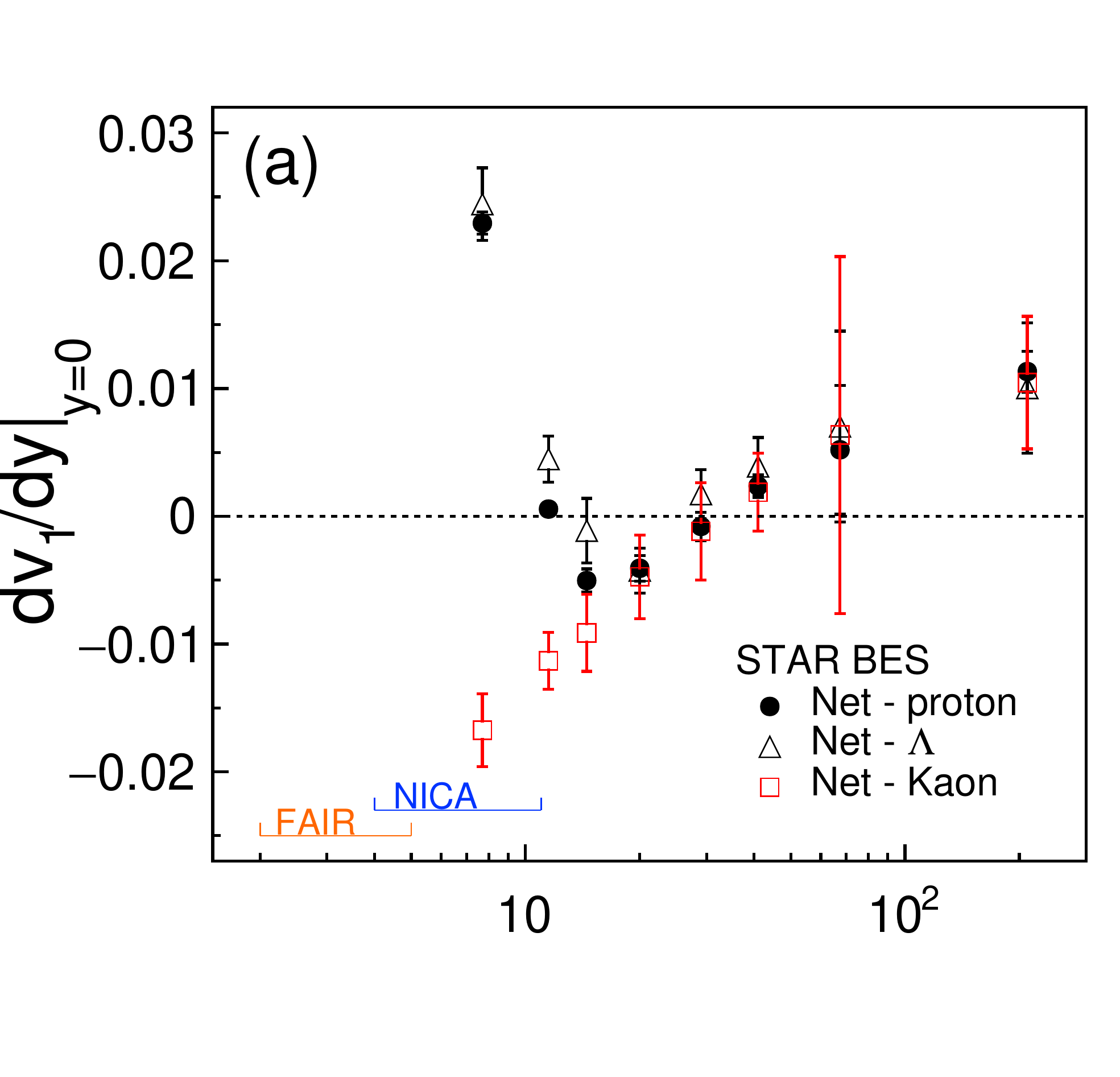}
\label{fig:side:a}
\end{minipage}
\hfill
\begin{minipage}[t]{0.5\linewidth}
\centering
\includegraphics[width=0.95\textwidth]{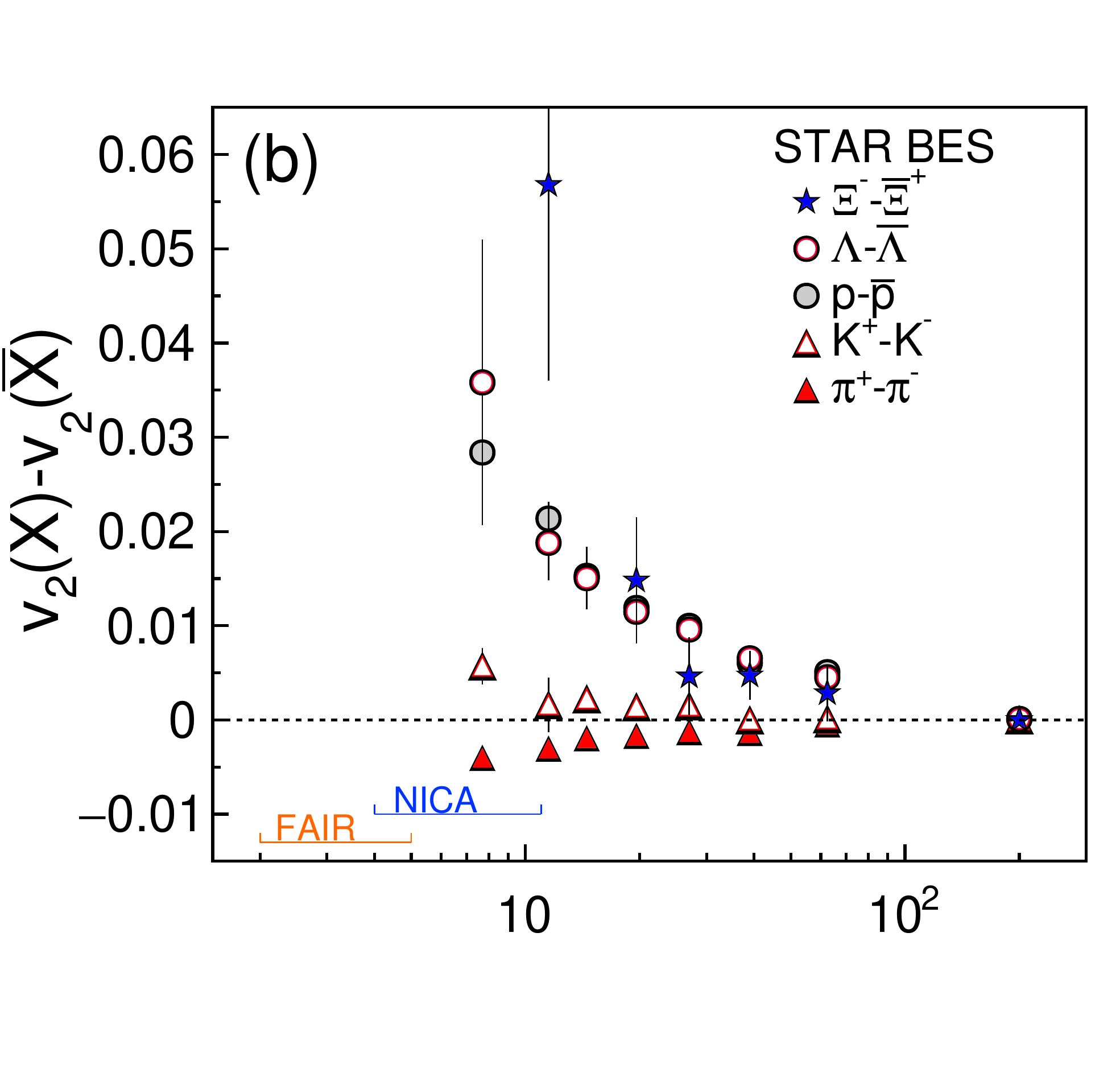}
\end{minipage}
\put(-320,10){\Large  \text { Collision Energy $\sqrt{\rm s_{NN}}$ (GeV)}}
\caption{(Color online) (\textbf{a}) The slope of $v_1$ at mid-rapidity ($dv_1/dy$) as a function of collision energy in 10\%--40\% Au+Au collisions for net-proton, net-$\Lambda$ and net-kaon~\cite{starv1_1, starv1_2}. 
(\textbf{b})The $v_2$ difference between particles and the corresponding anti-particles as a function of collision energy~\cite{BESpidv2_1, BESpidv2_2, BESpidv2_3}. The~energy regions of future collider and fix-target experiments, NICA and FAIR, are indicated in both plots.}
\label{BESflow}
\end{figure}

The first order coefficient of final azimuth distribution in the momentum space, $v_1$ (rapidity-odd), as~a function of rapidity is sensitive to   
the system expansion during the early stage of collisions.
Both hydrodynamic and nuclear transport models indicate that $v_1$ in the midrapidity region offers sensitivity to details of the expansion of the participant matter during the early collision stages~\cite{directedflow1, directedflow2, directedflow3}. 
Hydrodynamic plus first-order phase transition calculations suggest a minimum of net-baryon $v_1$ slope ($dv_1/dy$) near mid-rapidity is a signal of
phase transition between QGP and hadronic matter~\cite{v1phasetransition1, v1phasetransition2}. Net-particle is defined as the excess yield of a particle type
over its anti-particle~\cite{starv1_1, starv1_2}. The~$v_1$ of net-particle is defined as:
$v_{1X} = r(y)v_{1\bar{X}} + [1-r(y)]v_{1 \rm{net}-X}$, where $X$ represents particle, $\bar{X}$ represents the corresponding anti-particle,
$r(y)$ is the ratio of particle to anti-particle yield. Plot (a) of Figure~\ref{BESflow} shows the $v_1$ slope relative to rapidity for net-proton, net-$\Lambda$ and
net-kaon. Similar energy dependence is observed for net-proton and net-$\Lambda$. The~non-monotonic behavior is consistent with the hydrodynamical
calculations with first-order phase transition~\cite{v1phasetransition1, v1phasetransition2}.  Large divergence between $dv_1/dy$ of net-kaon and 
net-proton (net-$\Lambda$) is observed below $\sqrt{s_{NN}} < 20$ GeV, whereas all three agrees well at and above 20 GeV.
More theoretical inputs are needed to understand the difference. 
At the same time the measurements of centrality dependence in the future BES program will further verify the energy dependence and constrain model calculations.
In Ref.~\cite{starv1_2}, the~$dv_1/dy$ of $\phi$ mesons shows larger magnitude than pions and kaons at and above 14.5 GeV, 
and more interesting, the~$\phi$ meson slope seems to increase sharply at 11.5 GeV. 
Because of the large statistical uncertainties, it is still not conclusive. 
It~opens a new direction for both experimental and theoretical investigation on directed flow~\cite{v1ampt}.

As discussed above, the~$v_2$ of multi-strange hadrons and $\phi$ mesons are more sensitive to the parton level collectivity
as their hadronic cross sections are smaller than light flavor hadrons~\cite{multistrange1, multistrange2, multistrange_rev}. 
The~results from RHIC BES I suggest a possible drop of $\phi$ meson
$v_2$ compared to other hadrons in Au+Au collisions at $\sqrt{s_{NN}}$ = 11.5 and 7.7 with $\sim$2$\sigma$ effect~\cite{BESpidv2_1, BESpidv2_2, BESpidv2_3}. 
Data of high precision will be available with RHIC BES II.  
A significant difference in the $v_2$ values between particles and the corresponding anti-particles is observed at low energy heavy-ion collisions at RHIC.
As shown in plot (b) of Figure~\ref{BESflow}, the~difference is more pronounced for $v_2$ of baryons and anti-baryons when the collision energy 
is less than 20 GeV~\cite{BESpidv2_1, BESpidv2_2, BESpidv2_3}.
These differences naturally break the $n_q$ scaling discussed previously, as~the number of constituent quarks are same for particle and the corresponding anti-particle.
Several models try to explain the data~\cite{hybridv2, analytichydro, NJL1, NJL2, urqmdv2}: the hydro + transport (UrQMD) calculation can reproduce the proton data, but~not the meson data~\cite{hybridv2};
A analytic hydro model can quantitatively reproduce the $\pi$, $K$ and proton data, but~the flavor dependence 
($\Delta v_2^p > \Delta v_2^\Lambda > \Delta v_2^\Xi > \Delta v_2^\Omega$) is not consistent with data~\cite{analytichydro};
A Nambu-Jona-Lasino (NJL) model incorporating partonic and hadronic potentials can describe the data qualitatively, but~not quantitatively~\cite{NJL1, NJL2}.
New data from RHIC BES II, especially data of multi-strange hadrons, could offer more constrains on the model~calculation.


\section{Beam Energy Dependence of the Higher-Order Cumulants of Net-Particle Multiplicity Distributions and Light Nuclei~Productions}
Fluctuations of conserved quantities, such as net-baryon ($B$), net-charge ($Q$) and net-strangeness ($S$), are sensitive observables to search for the QCD critical point in heavy-ion collisions~\cite{Ejiri:2005wq,Stephanov:2008qz,Asakawa:2009aj,Gupta:2011wh,Luo:2017faz}. The~higher-order cumulants ($C_n$, the~$n^{\rm th}$ order cumulants), which can be used to quantify the fluctuations and describe the shape of the event-by-event multiplicity distributions, are predicted to be sensitive to the correlation length ($\xi$) of the system as $C_4 \propto \xi^{7}$ and $C_3 \propto \xi^{4.5}$~\cite{Stephanov:2008qz,Stephanov:2011pb}. The~various order cumulants and cumulant ratios can be expressed in terms of moments as 
$C_2=\sigma^{2}$, $C_3=S \sigma^{3}$,  $C_4=\kappa \sigma^{4}$ and $C_2/C_1=\sigma^{2}/M$, $C_3/C_2=S \sigma$, $C_4/C_2=\kappa \sigma^{2}$, where $\sigma^{2}$, $S$ and $\kappa$ are variance, skewness and kurtosis, respectively~\cite{Kitazawa:2017ljq}. The~various order cumulants are extensive quantities and are proportional to the system volume, which is difficult to be measured in heavy-ion collisions. 
By taking the ratio between various order cumulants, the~system volume can be cancelled to the first order and are directly related to the ratios of the thermodynamic susceptibilities ($\chi$) as $C_m/C_n=\chi^{(m)}/\chi^{(n)}$~\cite{Ejiri:2005wq,Gupta:2011wh,Ding:2015ona}.

\begin{figure}[H]
\centering
\includegraphics[width=2.1in]{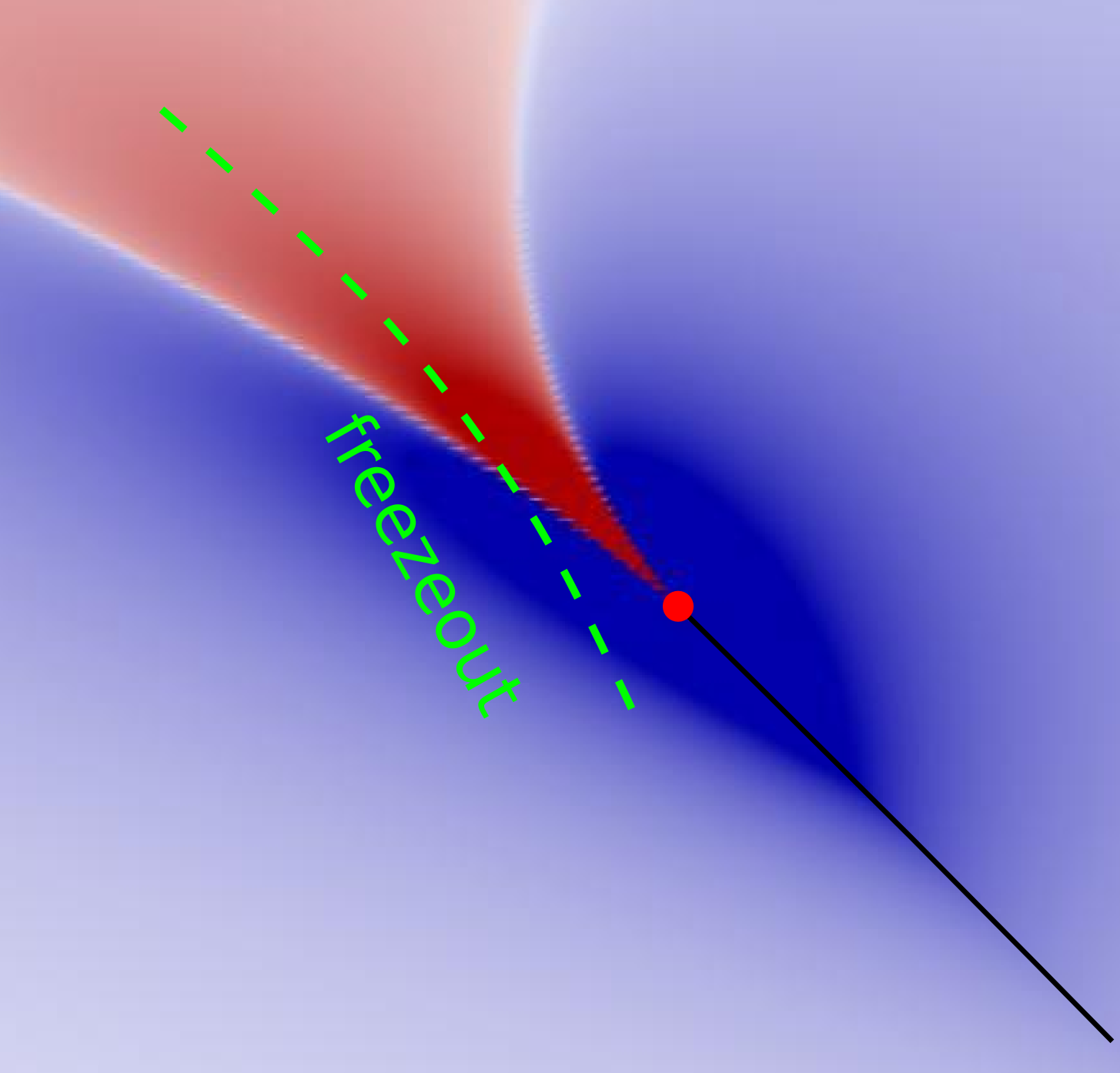}
\hspace{1cm}
\includegraphics[width=2.5in]{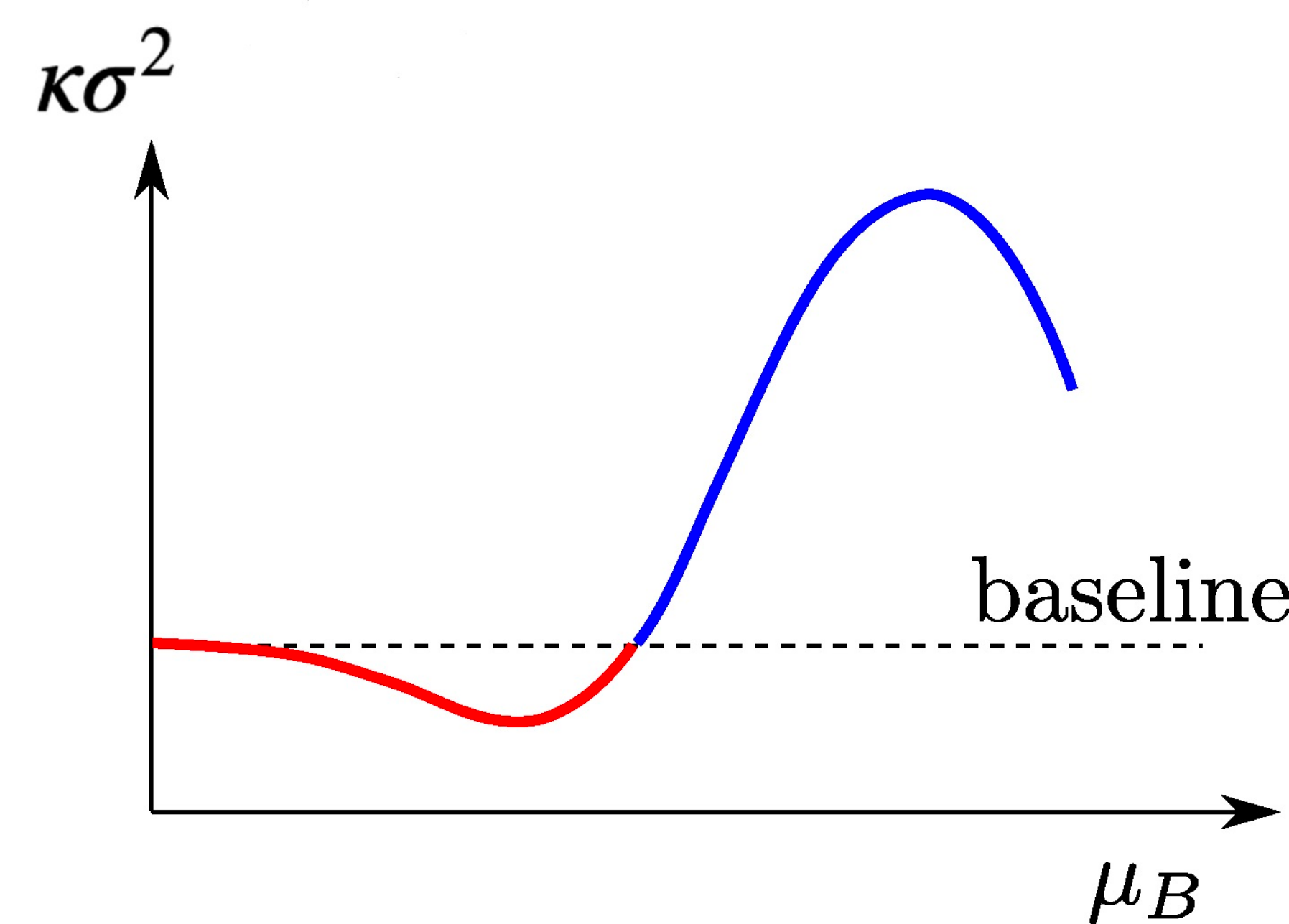}
\caption{(Color online) (Left) Density plot of fourth order cumulant of order parameter as a function of temperature and baryon chemical potential ($T$ and $\mu_B$) by mapping the Ising equation of state onto the QCD equation of state near the critical point~\cite{Bzdak:2019pkr}. The~red and blue regions in the density plot denote the negative and positive contributions to the fourth order cumulant, respectively.  The~green dashed line is the chemical freeze-out points ($T$, $\mu_B$) passing through the critical region when we scan the beam energies.  (Right) Normalized fourth order proton cumulant {\KV} as a function of collision energy or $\mu_B$ along the chemical freeze-out~line. } 
\label{fig:theory}
\end{figure}

Figure~\ref{fig:theory} (left) shows the density plot of fourth order cumulant of order parameter as a function of temperature and baryon chemical potential ($T$ and $\mu_B$) by mapping the Ising equation of state onto the QCD equation of state near the critical point~\cite{Bzdak:2019pkr}. The~red and blue regions in the density plot denote the negative and positive contributions to the fourth order cumulant, respectively. Experimentally, by~tuning the beam energy, the~$T$ and $\mu_B$ at chemical freeze-out are varied accordingly. The~green dashed line represents the chemical freeze-out points ($T$, $\mu_B$) passing through the critical region when one varies the beam energies. Figure~\ref{fig:theory} (right) shows fourth order fluctuation $\KV$ as a function of baryon chemical potential ($\mu_B$). Due to the negative and positive critical contributions near the critical point, the~$\KV$ will show a non-monotonic energy or $\mu_B$ dependence with respect to the non-critical baseline. This is the characteristic experimental signature of the critical point we are looking for in the heavy-ion collision experiment. Theoretically, the~properties of QCD phase diagram at finite baryon density and the signatures of conserved charge fluctuations near the QCD critical point have been extensively studied by various model calculations, such as Lattice QCD~\cite{Bazavov1,Bazavov:2012vg,Borsanyi:2014ewa,Bazavov:2015zja,Alba:2014eba,Bluhm:2018aei,Bazavov:2020bjn}, NJL, PNJL model~\cite{Fu:2009wy,Fu:2010ay,Lu:2015naa,Chen:2015dra,Fan:2016ovc,Fan:2017kym,Li:2018ygx,Yang:2019lyn}, PQM, FRG model~\cite{Friman:2011pf,Fu:2016tey,Fu:2019hdw}, Dyson-Schwinger Equation (DSE) method~\cite{Fischer:2012vc,Shi:2014zpa,Gao:2016qkh,Fischer:2018sdj}, chiral hydrodynamics~\cite{Herold:2016uvv} and other effective models~\cite{Stephanov:2011pb,Chen:2014ufa,Vovchenko:2015pya,Jiang:2015hri,Mukherjee:2016nhb,Zhang:2017icm}. However, one should keep in mind that the above results are under the assumption of thermal equilibrium with infinite and static medium. In~the real heavy-ion collisions, there exists the effects of finite size/time~\cite{Palhares:2010zz,Fraga:2011hi,Pan:2016ecs}, non-equilibrium~\cite{Mukherjee:2016kyu,Bluhm:2020mpc,Wu:2018twy,Nahrgang:2018afz,Asakawa:2019kek} and thermal blurring effects~\cite{Ohnishi:2016bdf}. Dynamical modeling of heavy-ion collisions by implementing both the critical and those background effects are ongoing~\cite{Stephanov:2017ghc,Rajagopal:2019xwg,An:2019csj}.  

Before turning to the experimental status, we would like to stress that there is a long history of using the higher-order cumulants to extract the information on that state created in high-energy nuclear collisions. As~extensively discussed in Ref.~\cite{alicedata2}, hadron yields or their ratios, which are the first order moment of the multiplicity distributions, have been used for determine the nature of thermalization in such collisions. Once the thermalization is established~\cite{bedangaetal19}, on~the other hand, the~higher-orders cumulants of the very same distributions can be used to study the fine structures of the QCD matter. For~example, the~critical point~\cite{Stephanov:2011pb}, the~nature of the crossover transition~\cite{Aoki,Friman:2011pf} and the phase boundary~\cite{A.Bzdak18} at vanishing and large net-baryon region, respectively.

\begin{figure}[H]
\centering
  \includegraphics[scale=0.62]{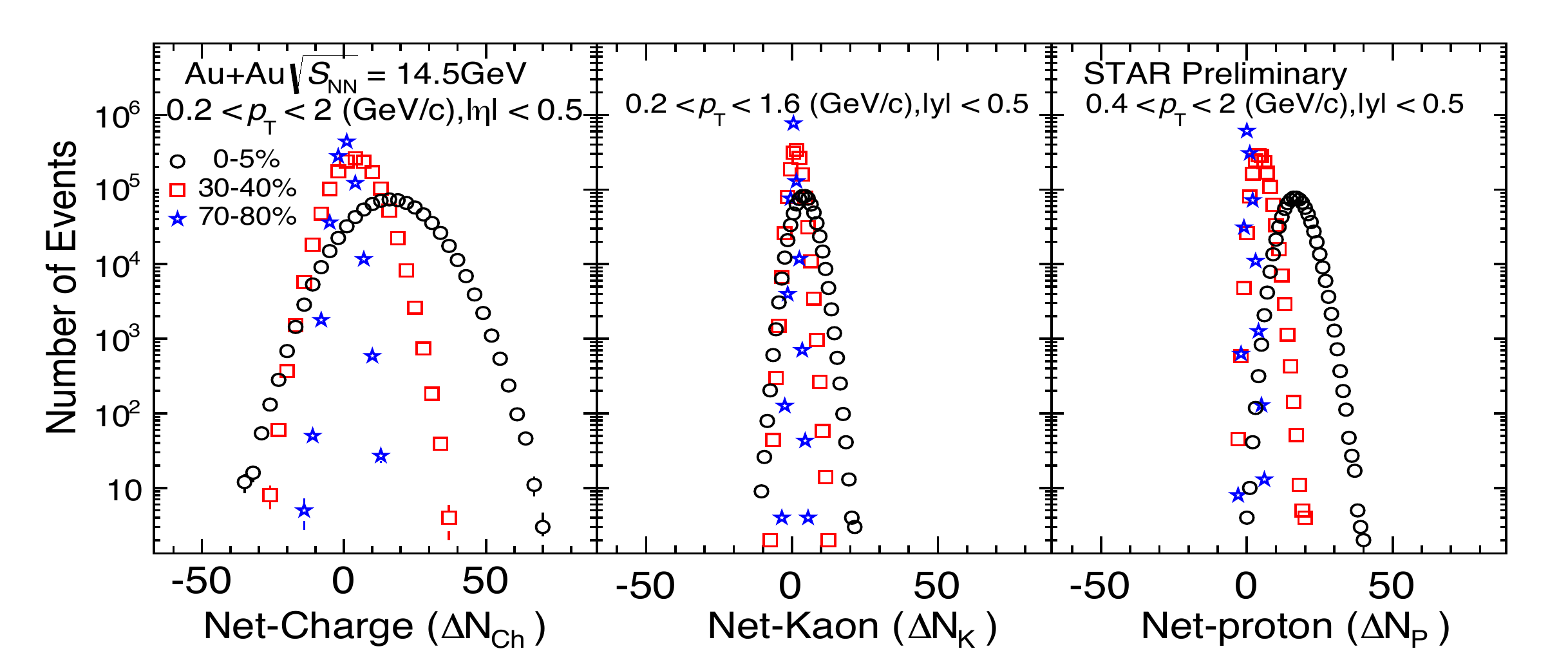}
       \caption{\textls[-20]{(Color online) The STAR 
       measured raw event-by-event net-charge, net-kaon and net-proton distributions of three centralities (0--5\%, 30\%--40\% and 70\%--80\%) in Au+Au collisions at {\sNN} = 14.5~GeV~\cite{Thader:2016gpa}.}} \label{fig:distribution}
\end{figure}

Experimentally, the~fluctuation of the net-proton and net-kaon are used as a proxy of net-baryon and net-strangeness fluctuations, respectively. The~STAR experiment has measured the higher-order cumulants ($C_1$--$C_4$) 
and second-order off-diagonal cumulants of net-proton~\cite{Aggarwal:2010wy,Adamczyk:2013dal,Luo:2015ewa,Luo:2015doi,Adam:2020unf,Adam:2019xmk}, net-charge~\cite{Adamczyk:2014fia} and net-kaon~\cite{,Adamczyk:2017wsl} multiplicity distributions in Au+Au collisions at $\sqrt{s_\mathrm{NN}}$= 7.7, 11.5, 14.5, 19.6, 27, 39, 62.4 and 200 GeV, which are collected during the first phase of RHIC beam energy scan program (2010--2014)~\cite{Aggarwal:2010cw}. To~make precise measurements, various corrections and techniques have been applied in the data analysis, those include : \textls[-5]{(1)~Select proper collision centralities to avoid auto-correlations and suppress volume fluctuations~\cite{Luo:2013bmi,Chatterjee:2019fey}, (2)~Detector Efficiency Correction~\mbox{\cite{Bzdak:2013pha,Luo:2014rea,Kitazawa:2016awu,Nonaka:2017kko,Luo:2018ofd}}}, (3)~Centrality Bin Width Correction (CBWC)~\cite{Luo:2013bmi}, (4)~Statistical error estimation with Delta theorem and/or Bootstrap~\cite{Luo:2011tp,Luo:2013bmi}. Figure~\ref{fig:distribution} shows the measured event-by-event net-charge, net-kaon and net-proton multiplicity distributions of three different centralities in Au+Au collisions at {\sNN} = 14.5 GeV. Those are raw distributions and not corrected for detector efficiency and acceptance. One should apply the efficiency correction and CBWC to obtain the final efficiency corrected cumulants. In~Ref.~\cite{Luo:2011tp}, we have shown that the statistical uncertainties of the cumulants ($C_n$) strongly depend on the width of the distributions ($\mathrm{error} \propto \sigma^{n}/\sqrt{\mathrm{N}}$). In~general, the~widths of the distributions in central collisions are wider and with larger mean values than those from peripheral collisions. Further, the~widths of the net-charge distributions are much wider than those of net-proton and net-kaon in the same centrality. That's the reason why we observe larger statistical uncertainties in central collisions than those from peripheral. Assuming the measured particles are emitted from many independent sources in the fire ball, the~multiplicity distributions in central collisions would be more symmetric and close to Gaussian distribution than the peripheral based on the central limit theorem (CLT). If~those sources are identical and uncorrelated, the~cumulant ratios are expected to be a constant as a function of collision~centralities. 

\begin{figure}[H]
\centering
\includegraphics[scale=0.75]{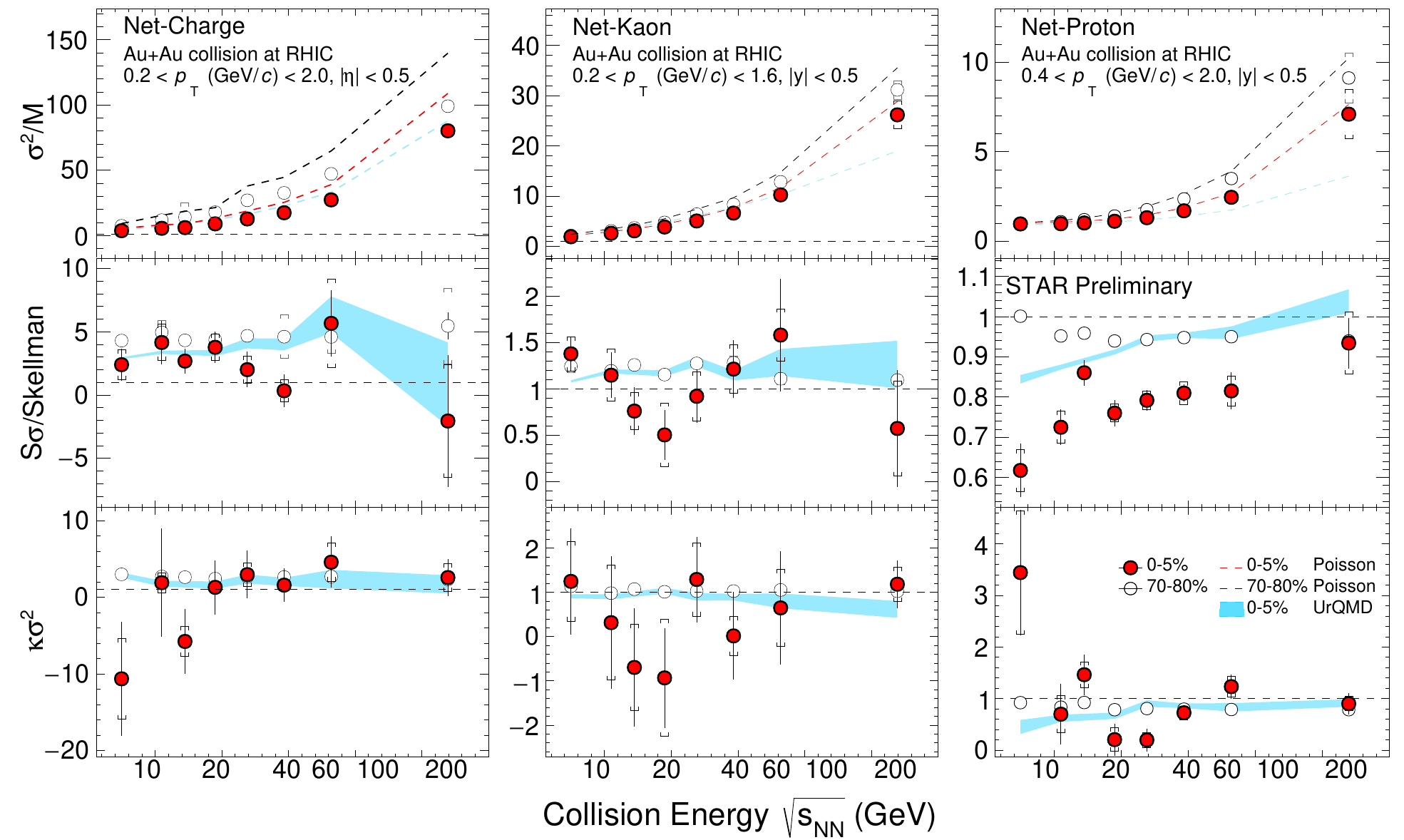}
\caption{(Color online) Energy dependence of cumulant ratios ($\sigma^{2}/M$, $S\sigma/$Skellam, $\kappa\sigma^2$) of net-charge, \textls[-15]{net-kaon and net-proton multiplicity distributions for top 0--5\% and~70\%--80\% peripheral collisions. The~Poisson expectations are denoted as dotted lines and UrQMD 
calculations are shown as bands. The~statistical and systematical errors are shown in bars and brackets,~respectively.}} 
\label{fig:egyDependence}
\end{figure}

Figure~\ref{fig:egyDependence} shows the energy dependence of cumulant ratios ($\sigma^{2}/M$, $S\sigma/$Skellam, $\kappa\sigma^2$) of net-charge, net-kaon and net-proton multiplicity distributions in Au+Au collisions measured by STAR. The~blue bands are the results obtained from UrQMD model calculations without including the physics of critical point~\cite{Xu:2016qjd,Zhou:2017jfk,Luo:2017faz}. The~{\SD} values are normalized by the Skellam expectations, which are constructed with the measured mean values of proton and anti-proton by assuming they are distributed as independent Poisson distributions. The~deviation of {\SD}/skellam from unity would indicate the deviation of {\SD} from Poisson statistical fluctuations (Poisson baseline). For~{\SD}/Skellam and {\KV}, their Poisson baselines are unity, which are plotted as the dashed lines. We found that the {\VM} of net-charge, net-kaon and net-proton monotonically increase when increasing the collision energy. The~$S\sigma/$Skellam and $\kappa\sigma^2$ show weak energy dependence for net-charge and net-kaon measurements. We didn't observe significant deviations of net-charge and net-kaon cumulant ratios $S\sigma/$Skellam and {\KV} from the Poisson expectations and UrQMD calculations within uncertainties. However, a~clear non-monotonic energy dependence of net-proton $\kappa\sigma^2$ was observed in top 0--5\% central Au+Au collisions. The~0--5\% net-proton {\KV} values are close to unity for energies above 39 GeV and show large deviations below unity around 19.6 and 27 GeV, and~then increasing above unity below 19.6~GeV. The~UrQMD calculations of net-proton {\KV} displaying a strong suppression below unity at lower energies is due to the effects of baryon number conservation~\cite{Bzdak:2012an,Braun-Munzinger:2016yjz,He:2016uei,He:2017zpg}. However, this suppression is not observed at low energies in the STAR data. Another transport model (JAM model) study further demonstrates that the resonance weak decay and hadronic re-scattering have very small effects on the proton number fluctuations ($C_1$--$C_4$) at low energies~\cite{Zhang:2019lqz}. 

\begin{figure}[H]
\centering
    \includegraphics[scale=0.7]{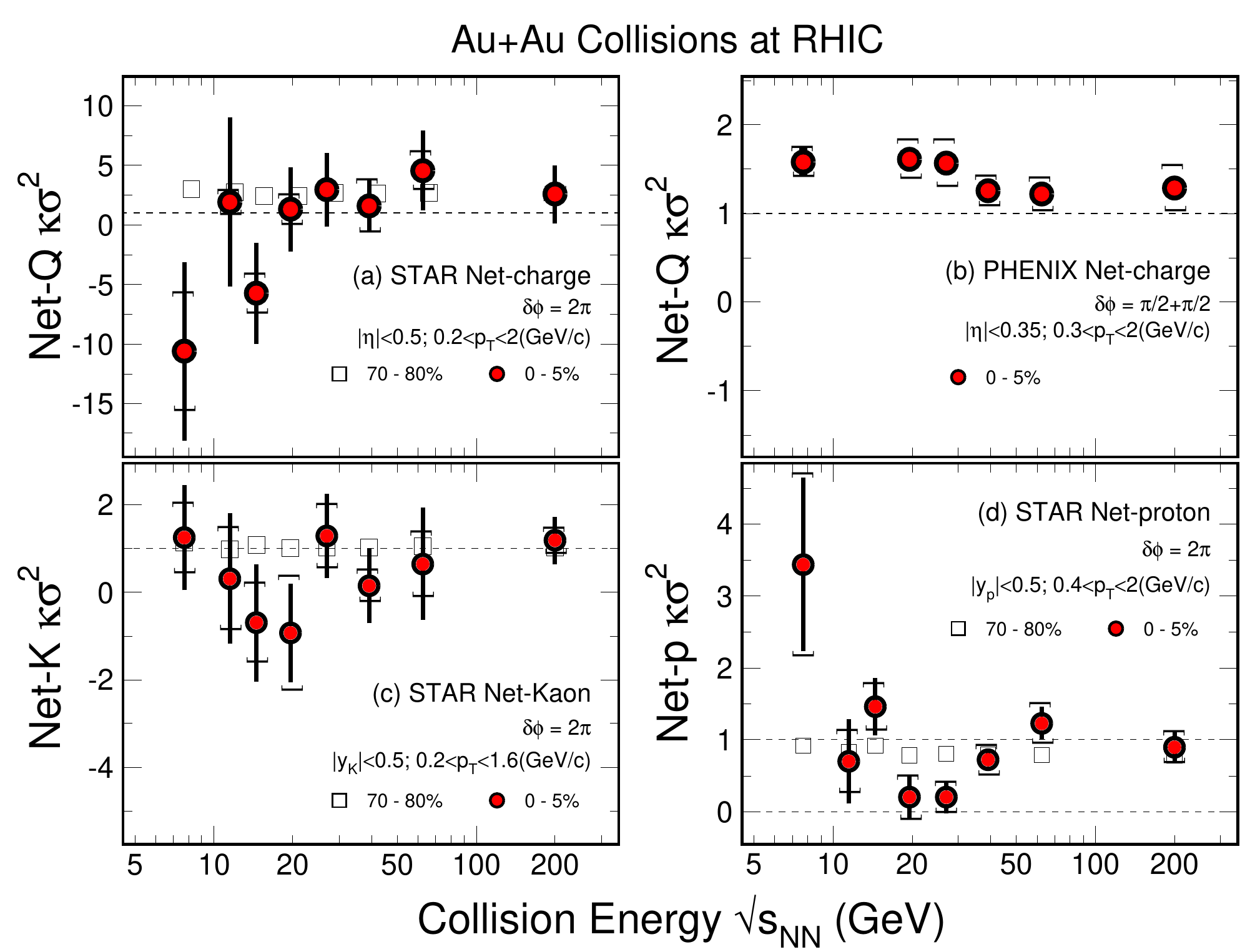},
       \caption{(Color online) The STAR measured energy dependence of $\kappa\sigma^2$ of net-charge (top left), net-kaon and net-proton distributions in Au+Au collisions at {\sNN}= 7.7, 11.5, 14.5, 19.6, 27, 39, 62.4 and 200~GeV. The~net-charge fluctuations measured by the PHENIX experiment in Au+Au collisions at \mbox{{\sNN}= 7.7}, 19.6, 27, 39, 62.4 and 200 GeV are shown in top right panel.The statistical and systematical errors are shown in bars and brackets, respectively.} \label{fig:KV_energy}
\end{figure}

In Figure~\ref{fig:KV_energy}, we summarize the energy dependence of {\KV} of net-charge, net-kaon and net-proton multiplicity distributions in Au+Au collisions measured by the STAR experiment. For~comparison, the~net-charge results in Au+Au collisions at {\sNN}= 7.7, 19.6, 27, 39, 62.4 and 200 GeV measured by the PHENIX experiment~\cite{Adare:2015aqk} are shown in the panel (b). We found that the {\KV} of the net-charge and net-kaon multiplicity distributions measured by the STAR experiment show larger statistical uncertainties than those of net-proton {\KV}. This can be understood as the statistical uncertainties of {\KV} depend on the width ($\sigma$) of the multiplicity distributions and the particle detecting efficiencies ($\epsilon$) in the detector as $error(C_n/C_2) \propto \sigma^{n-2}/(\sqrt{N}  \epsilon^{n})$~\cite{Luo:2014rea}. The~width of the net-charge distributions are larger than those of net-proton and net-kaon. Meanwhile, due to decays, the~efficiency of kaon ($\sim$40\%) is much lower than proton ($\sim$80\%). It is the reason why we observe larger statistical uncertainties for net-kaon fluctuations than net-proton. For~the net-charge and net-kaon {\KV} from STAR, we observe weak energy dependence within current statistical uncertainties. The~PHENIX net-charge {\KV} are with much smaller statistical uncertainties than the results from STAR. This is due to smaller acceptance of PHENIX detector than the STAR detector, thus the width of the net-charge multiplicity distributions measured by the PHENIX experiment is much narrower than those measured by STAR. We observe a clear non-monotonic energy dependence for net-proton {\KV} in the most central (0--5\%) Au+Au collisions with a minimum around 19.6~GeV. This non-monotonic behavior cannot be described by various model calculations without the physics of phase transition and critical~point. 

\begin{figure}[H] 
\centering
	\includegraphics[width=4.5in]{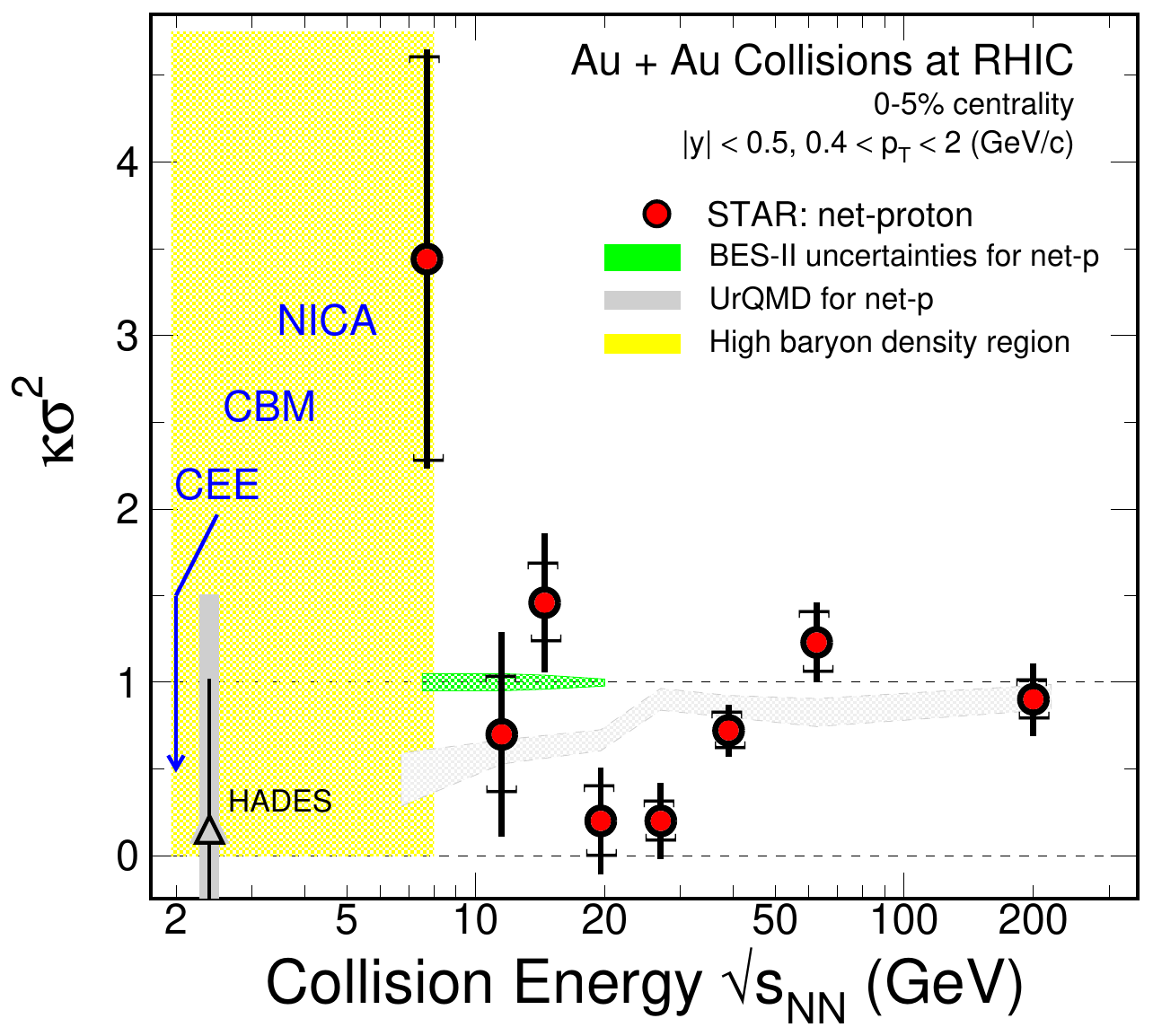}
	\caption{(Color online) Energy dependence of the mid-rapidity net-proton 4th order cumulants ratios from central 0--5\% Au+Au collisions, STAR experiment ($|y|< 0.5, 0.4 < p_T < 2.0$ GeV/c: filled-circles)~\cite{Adam:2020unf} and 0--10\% Au+Au collision, HADES experiment ($|y| < 0.4,  0.4 < p_T < 1.6$ GeV/c: open-triangle)~\cite{HADES:2020}. The statistical uncertainties of the second RHIC beam-energy-scan (BES-II) are shown as the green-band while the UrQMD results are shown as gray-band. The~energy region covered by the future experiments is shown as~yellow.} \label{fig:KV_CBM}
\end{figure}

\textls[-5]{Figure~\ref{fig:KV_CBM} shows the energy dependence of the fourth-order fluctuations ({\KV}) of net-proton from the top 5\% central Au+Au collisions measured by STAR experiment~\cite{Adam:2020unf}. Recent result from the HADES experiment is also shown in the figure. Note that there are differences in the data shown in Figure~\ref{fig:KV_CBM} : while STAR data points are from the top 5\% central collisions and $|y| < 0.5, 0.4 < p_T < 2.0$~GeV/c}, the~HADES data is from the top 10\% central Au+Au collisions and $|y| < 0.4, 0.4 < p_T < 1.6$~GeV/c. From~200 GeV to 7.7 GeV, non-monotonic energy dependence is clearly shown in the {\KV} of net-proton multiplicity distributions and one can observe a strong enhancement at the highest  $ \mu_B \sim 420$ MeV, corresponding to the Au+Au central collisions at \sNN\ = 7.7 GeV. This might indicate attractive correlations between nucleons in nature at the large baryon density region. However, interestingly, the~strong enhancement in the fourth-order fluctuation seems disappeared as shown by the HADES result at \sNN\ = 2.4 GeV~\cite{HADES:2020}. Indeed, in~the high baryon density region, between~\sNN\ = 2 GeV and 8 GeV, there  might be a peak in the fourth order fluctuations as speculated in Ref.~\cite{Luo:2017faz,A.Bzdak18,Bzdak:2019pkr}. If~the peak structure is confirmed, that would be the experimental indication of the QCD critical point and/or the first order phase transition created in such high-energy nuclear collisions. On~the other hand, it is possible that from \sNN\ = 2 GeV to 8 GeV data points are smoothly connected without any peaks or dips. Results from the future experiments like NICA, CBM, and~CEE will certainly provide the~answer.     

In Figure~\ref{fig:KV_CBM}, the~results from the transport model UrQMD (grey band) show a monotonic decrease from low to high baryon density region, which is due to the effect of baryon number conservation in high-energy nuclear collisions. Note that in the Poisson limit, the~absence of criticality or other dynamical correlations, the~{\KV} is expected to be unity. The~green band in the figure is the projected statistical error of the fourth-order fluctuations {\KV} of net-protons in the second phase of the RHIC Beam Energy Scan (BES-II, 2019--2021) program~\cite{BESII_WhitePaper}. The~BES-II program, which is scheduled between 2019 and 2021 for the Au+Au collisions at 7.7--19.6 GeV, will take about 10 to 20 times higher statistics (depending on energy) to confirm the non-monotonic behavior observed in the fourth order fluctuations ({\KV}) of net-proton and proton in Au+Au collisions in the BES-I at RHIC. Assuming the data in the figure is related to the critical region, one must study the net-proton fluctuations at even higher baryon density region, i.e.,~at lower collision energies. The yellow band shown in the figure represents the high baryon
density region (\sNN\ = 2--8 GeV)
covered by future FAIR/CBM fixed target (FXT) experiment (\sNN\ =
2--5 GeV)~\cite{Ablyazimov:2017guv} and the NICA/MPD collider
experiment (\sNN = 4--11 GeV)~\cite{NICA_WhitePaper}.. 

\begin{figure}[H] 	
\hspace{2.8cm}
	\includegraphics[width=3.8in]{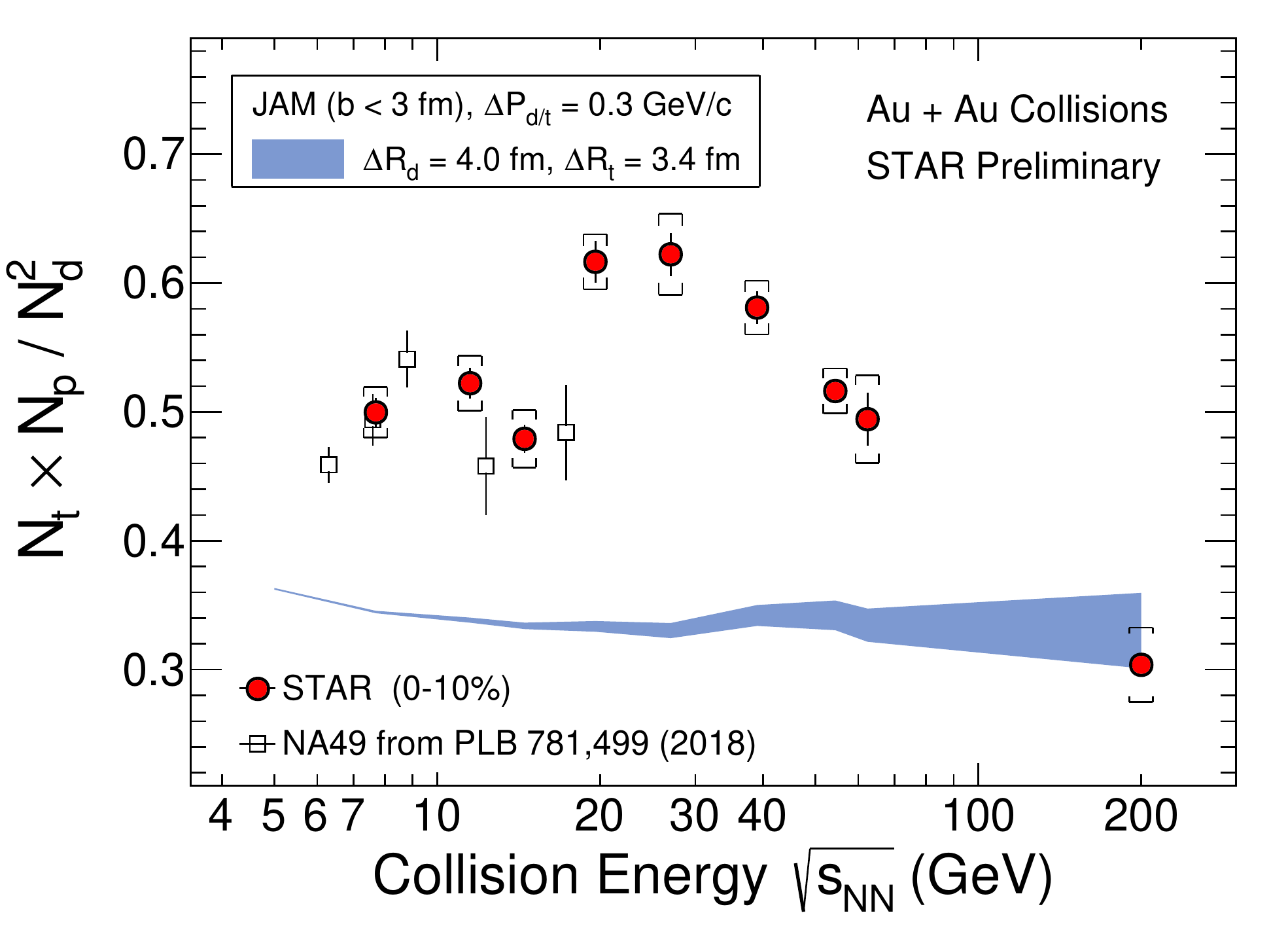}
	\caption{(Color online) Energy dependence of the light nuclei yield ratio $N_t \times N_p /N^{2}_d$ in central (0--10\%) heavy-ion collisions. The~red solid circles are the results measured in central (0--10\%) Au+Au collisions at BES energies by the STAR experiment and the open squares are the results calculated from the Pb+Pb data of NA49 experiment. The~blue band represents the results of central Au+Au collisions ($b<3$ fm) from the JAM model calculations~\cite{Liu:2019nii}.} \label{fig:light_nuclei}
\end{figure}

Besides the conserved charge fluctuations, the~light nuclei production is predicted to be sensitive to the baryon density fluctuations assuming that the light nuclei is formed from the nucleon coalescence. Model calculations show that the yield ratio between deuteron, triton and proton, $N_t \times N_p /N^{2}_d$ is related to the neutron density fluctuations, thus can be used to search for the QCD critical point in heavy-ion collisions~\cite{Sun:2018jhg, Yu:2018kvh}. Experimentally, the~STAR experiment has measured the production of deuteron ($d$) and triton ($t$) in the Au+Au collisions at {\sNN}= 7.7, 11.5, 14.5, 19.6, 27, 39, 54.4, 62.4 and 200 GeV. As~shown in Figure~\ref{fig:light_nuclei}, non-monotonic energy dependence is observed for the yield ratio, $N_t \times N_p /N^{2}_d$, in~0--10\% central Au+Au collisions with a peak around 20--30 GeV~\cite{Liu:2019nii, Zhang:2019wun, Zhang:QM2019}. The~yield ratios measured by STAR experiment below 20 GeV are consistent with the results calculated from NA49 experiment~\cite{Sun:2018jhg}. Since there is no critical physics implemented in the JAM model, the~results of central ($b<3$ fm) Au+Au collisions from JAM model is also plotted as blue band in Figure~\ref{fig:light_nuclei} for comparison~\cite{Liu:2019nii}. The~model results show a flat energy dependence and cannot describe the non-monotonic trend observed in the STAR data. The~current STAR results shown in Figure~\ref{fig:light_nuclei} is for 0--10\% centrality, it is also worthwhile to perform centrality dependence study on this yield ratio. On~the other hand, more theoretical studies and dynamical modeling of heavy-ion collisions with critical physics are needed to understand whether this non-monotonic behavior is related to the QCD critical~fluctuations.

\section{Beam Energy Dependence of the Heavy-Flavor~Production}

\begin{figure}[H]
\centering
\includegraphics[width=9 cm]{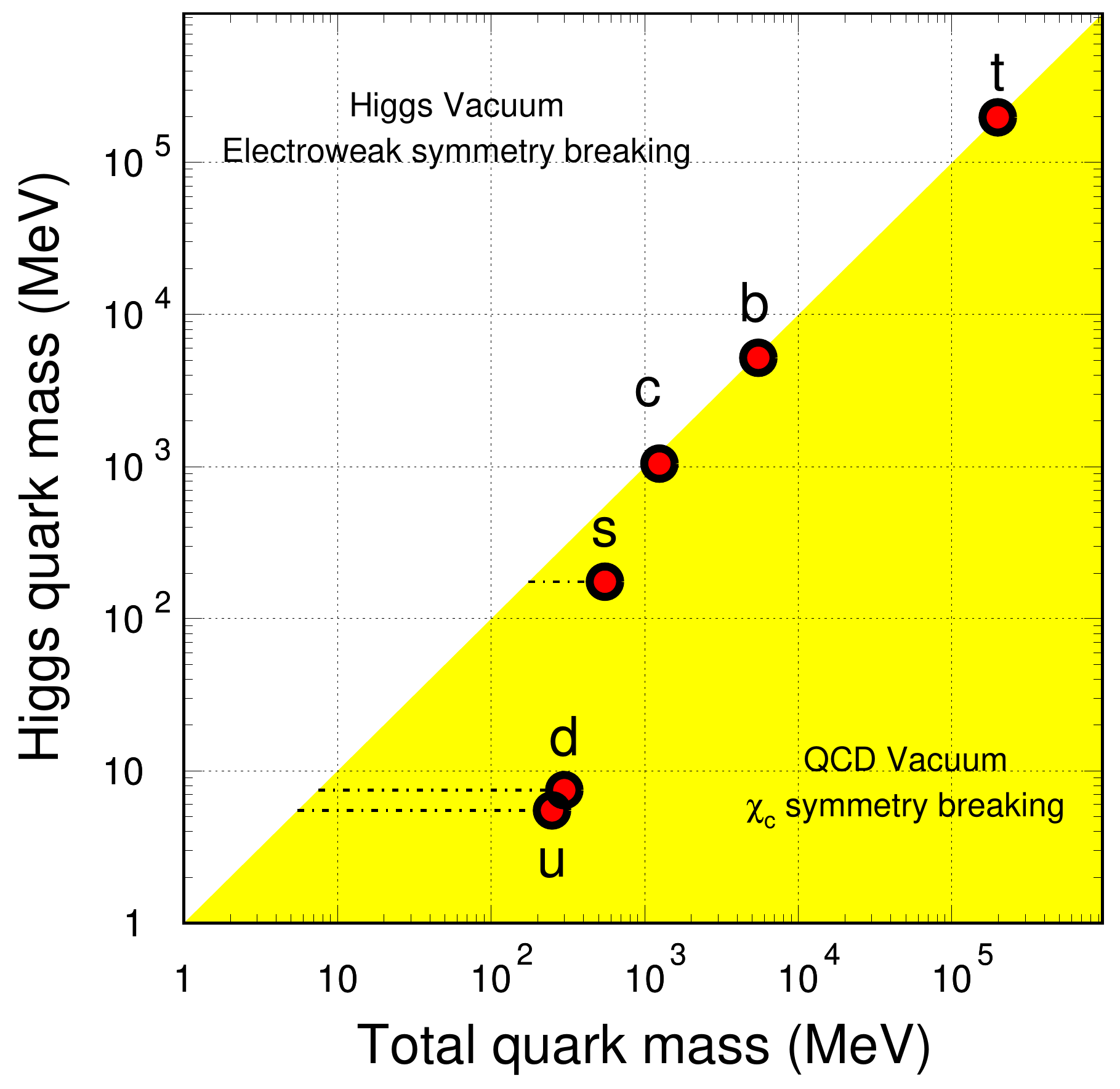}
\caption{(Color online) Quark masses in the QCD vacuum and the Higgs vacuum. A~large fraction of the light quark mass is due the the chiral symmetry breaking in the QCD vacuum. The~numerical values were taken from reference~\cite{hqmass1}. The~plot is taken from Ref.~\cite{hqmass2}.}
\label{hqmass}
\end{figure}

Since the masses of heavy flavor quarks are much larger than the temperature of the system created in the high-energy nuclear collisions, they can be used as clean probes of the medium properties at early stage of the collisions.  As~shown in Figure~\ref{hqmass}, heavy flavor quark masses are all generated in the electro-weak sector while light quarks ($u$, $d$, and~$s$) are dominated by the spontaneous breaking of chiral symmetry in QCD. Thus heavy quarks keep massive when participating in strong interactions. Due to their large masses, these heavy flavor quarks are primarily pair-created in initial hard pQCD processes. These facts making heavy quark hadrons are ideal for studying the medium effects including the thermalization of the system. One example has already discussed in previous section, see Figure~\ref{D0v2}.

\begin{figure}[H]
\centering
\includegraphics[width=7 cm]{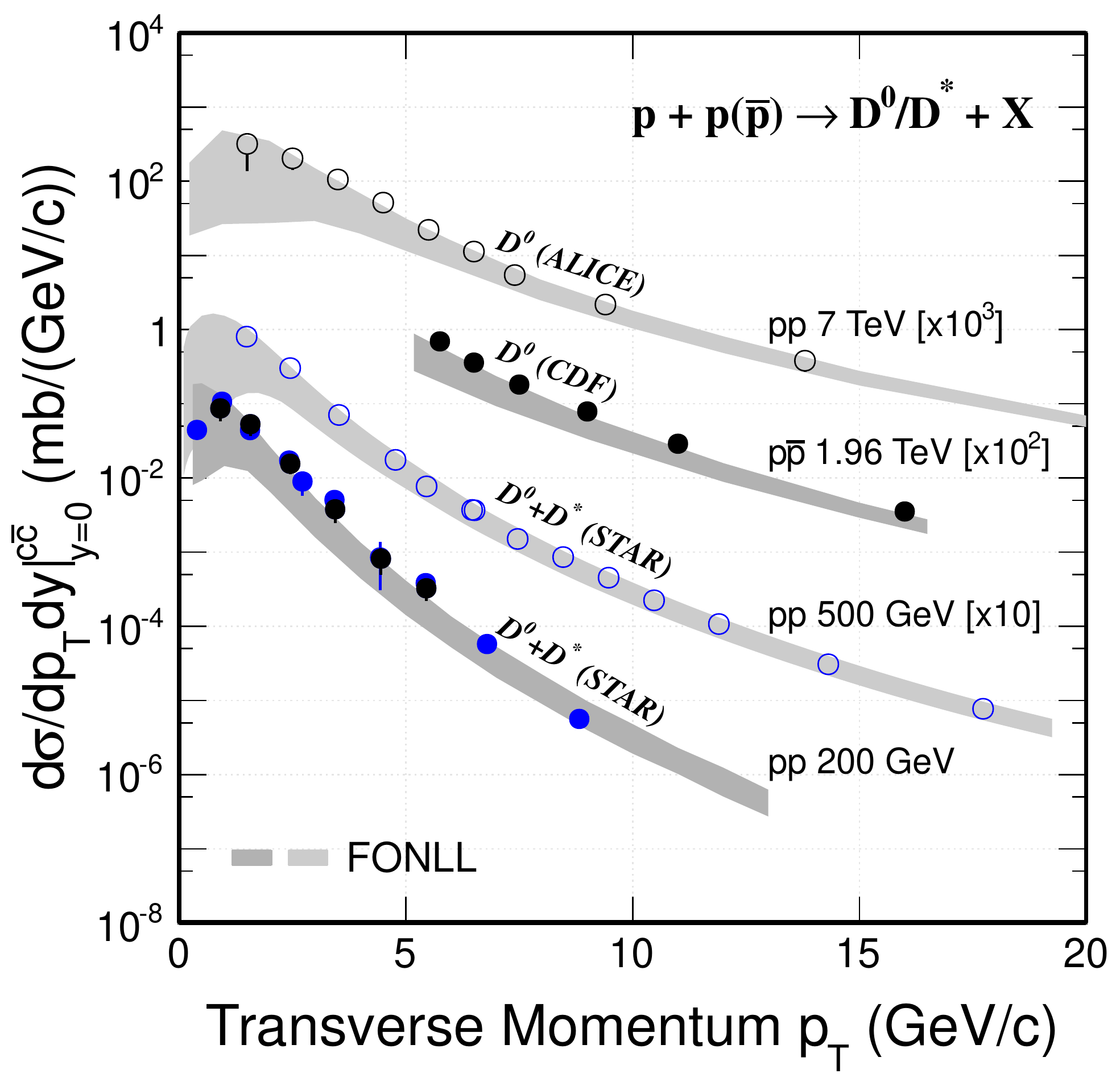}
\caption{(Color online) Charm production cross sections at midrapidity as a function of $p_{T}$. Symbols from top to bottom are experiment results at $\sqrt{s} = $ 7 TeV~\cite{ALICE12}, 1.96 TeV~\cite{CDF03}, 500 GeV~\cite{STARD500} and 200 GeV~\cite{STARDpp} from ALICE, CDF and STAR experiments, respectively. The~gray bands are FONLL calculations with~uncertainties.}
\label{cxsecpp}
\end{figure}

The QCD calculations can evaluate the charm production cross sections at high energies via a perturbation scheme in $p$+$p$ collisions~\cite{pQCD1, FONLL}. Figure~\ref{cxsecpp} shows the charm production cross sections at midrapidity as a function of $p_{\rm T}$ in $p$+$p$ collisions at $\sqrt{s} = $ 7 TeV, 1.96 TeV, 500 GeV and 200 GeV from ALICE~\cite{ALICE12}, CDF~\cite{CDF03} and STAR~\cite{STARDpp, STARD500} experiments, respectively. Within~uncertainties the Fixed-Order-Next-to-Leading-Logarithm (FONLL) calculations~\cite{FONLL} agree with data. The~data points are more on top of the upper limit of the theoretical uncertainties for all of the collision~energies.

\begin{figure}[H]
\centering
\includegraphics[width=12 cm]{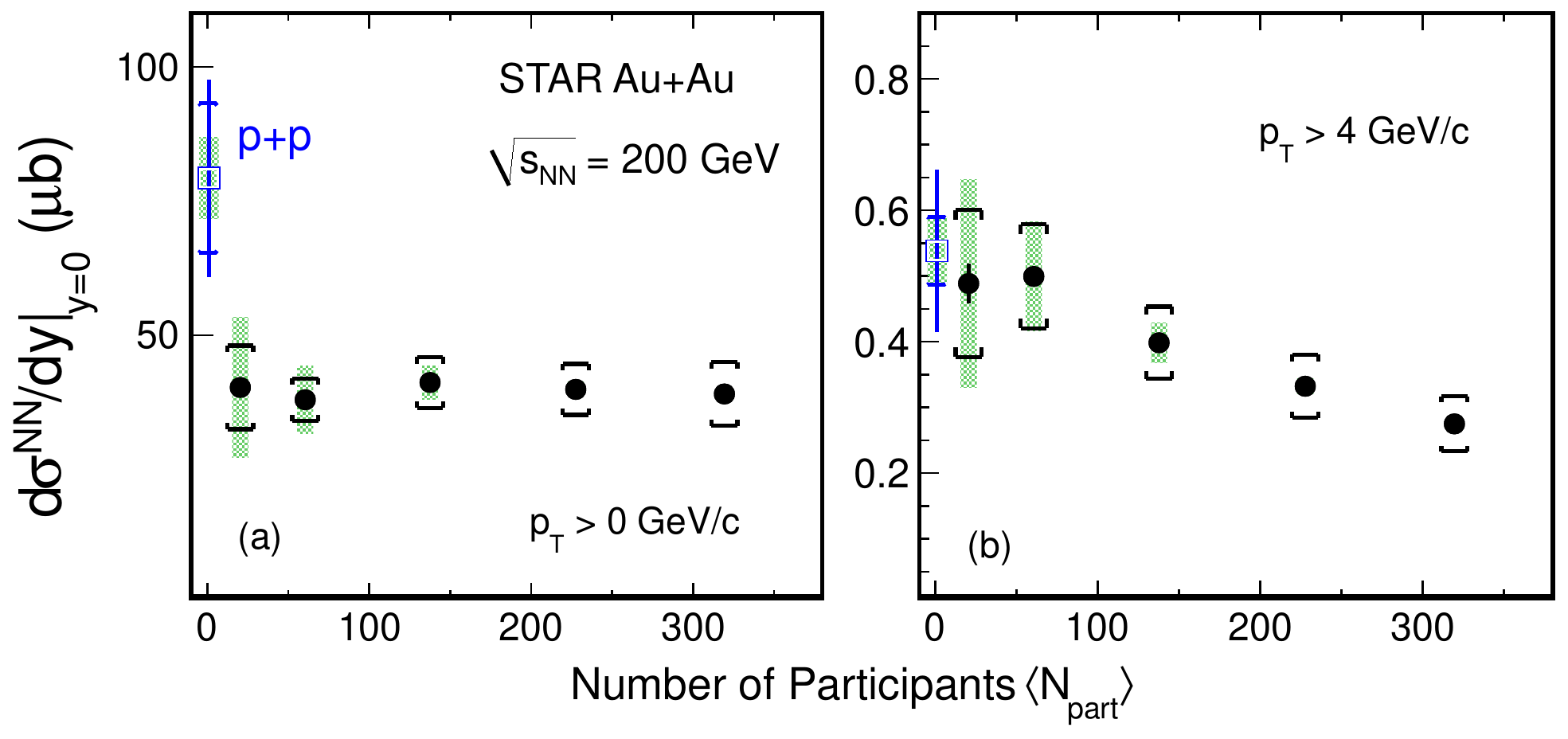}
\caption{(Color online) Integrated $D^0$ cross section per nucleon-nucleon collision at mid-rapidity in $\sqrt{s_{NN}} = $ 200 GeV Au+Au collisions for $p_{T}$ > 0 (\textbf{a}) and $p_{T}$ > 4 GeV/$c$ (\textbf{b}) as a function of centrality $N_{\rm part}$. The~statistical and systematic uncertainties are shown as bars and brackets on the data points. The~green boxes on the data points depict the overall normalization uncertainties in $p$+$p$ and Au+Au data~respectively. }
\label{cxsec}
\end{figure}

In heavy-ion collisions, charm quark interacts with the QGP matter when traversing in the medium. The~transverse momentum of charm quark is modified by the medium via energy loss or collective flow. \textls[-15]{However, the~total number of charm quarks may keep conserved since they are produced in initial hard processes before the QGP formation and there is no more charm quark created later via thermal production at RHIC energies. Figure~\ref{cxsec} shows the $p_{T}$-integrated cross section for $D^0$ production per nucleon-nucleon collision $d\sigma^{\rm NN}/dy|_{y=0}$ from different centrality bins in \mbox{$\sqrt{s_{NN}}$ =  200 GeV}} Au+Au collisions for the full $p_{T}$ range (a) and for $p_{T}$\,$>$\,4\,GeV/$c$ (b), respectively~\cite{STARDAA}. The~result from the $p$+$p$ measurement at the same collision energy is also shown in both panels~\cite{STARDpp}.

The high $p_{T}$ ($>$\,4\,GeV/$c$) $d\sigma^{\rm NN}/dy|_{y=0}$ shows a clear decreasing trend from peripheral to mid-central and central collisions and the result in peripheral collisions is consistent with $p$+$p$ collisions within uncertainties. This is consistent with charm quarks lose more energy in more central collisions at high $p_{T}$. However, for~the $d\sigma^{\rm NN}/dy|_{y=0}$ integrated over full $p_{T}$ range shows approximately a flat distribution as a function of $N_{\rm part}$. The~values for the full $p_{T}$ range in mid-central to central Au+Au collisions are smaller than that in $p$+$p$ collisions with $\sim$1.5$\sigma$ effect considering the large uncertainties from the $p$+$p$ measurements. The~total charm quark yield in heavy-ion collisions is expected to follow the number-of-binary-collision scaling since charm quarks are conserved at RHIC energies. However, the~cold nuclear matter (CNM) effect including shadowing could also play an important role. In~addition, hadronization through coalescence could alter the hadrochemistry distributions of charm quark in various charmed-hadron states which may lead to the reduction in the observed $D^0$ yields in Au+Au collisions~\cite{GRECO2004202}. For~instance, hadronization through coalescence can lead to an enhancement of the charmed baryon $\Lambda_{c}^+$ over $D^0$ yield~\cite{Oh2009,Zhao:2018jlw,Plumari:2017ntm}, and~together with the strangeness enhancement in the hot QCD medium and sequential hadronization, can also lead to an enhancement in the charmed strange meson $D_{s}^+$ yield relative to $D^0$~\cite{He2013,Zhao:2018jlw,Plumari:2017ntm}.

\begin{figure}[H]
\centering
\includegraphics[width=7 cm]{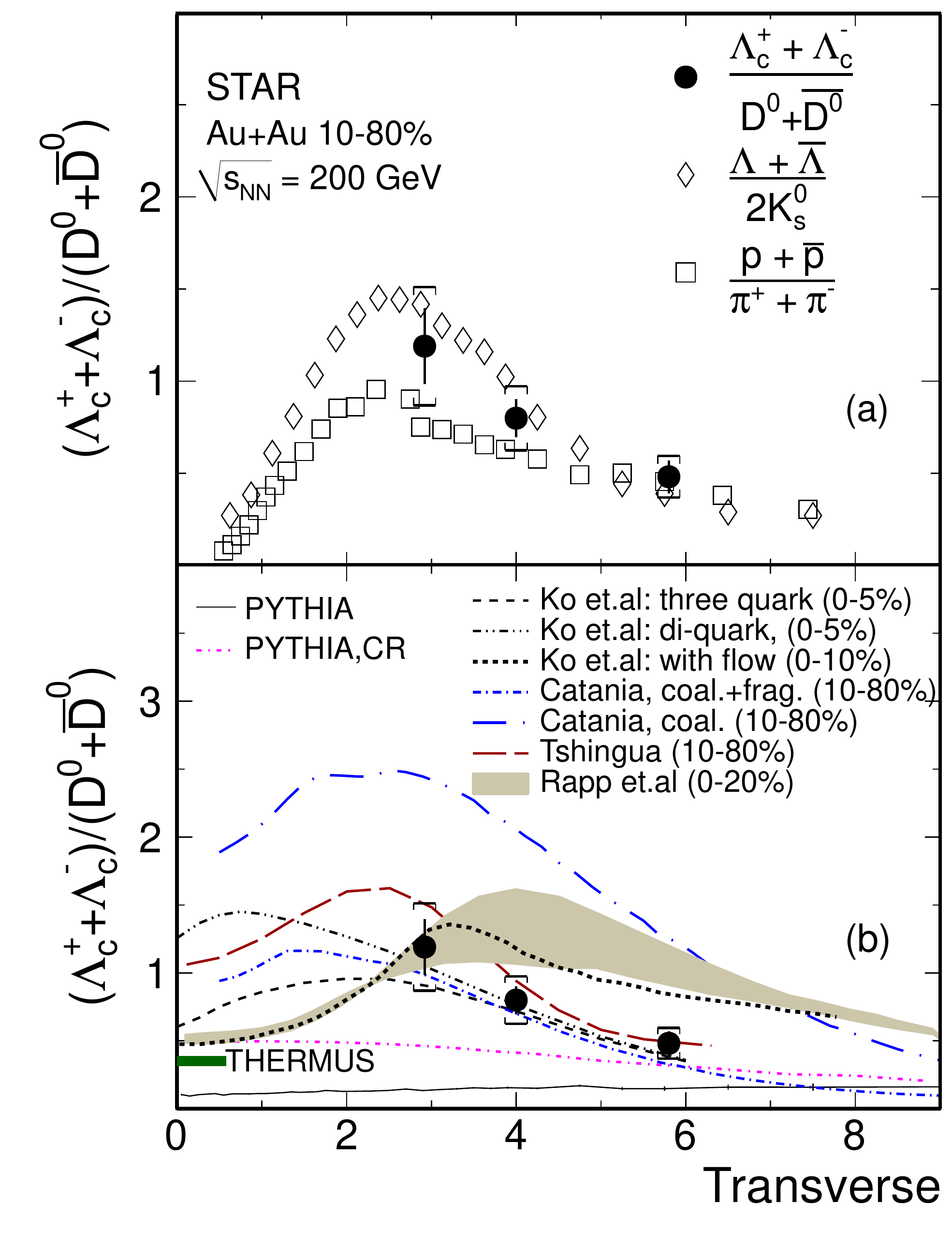}
\includegraphics[width=7 cm]{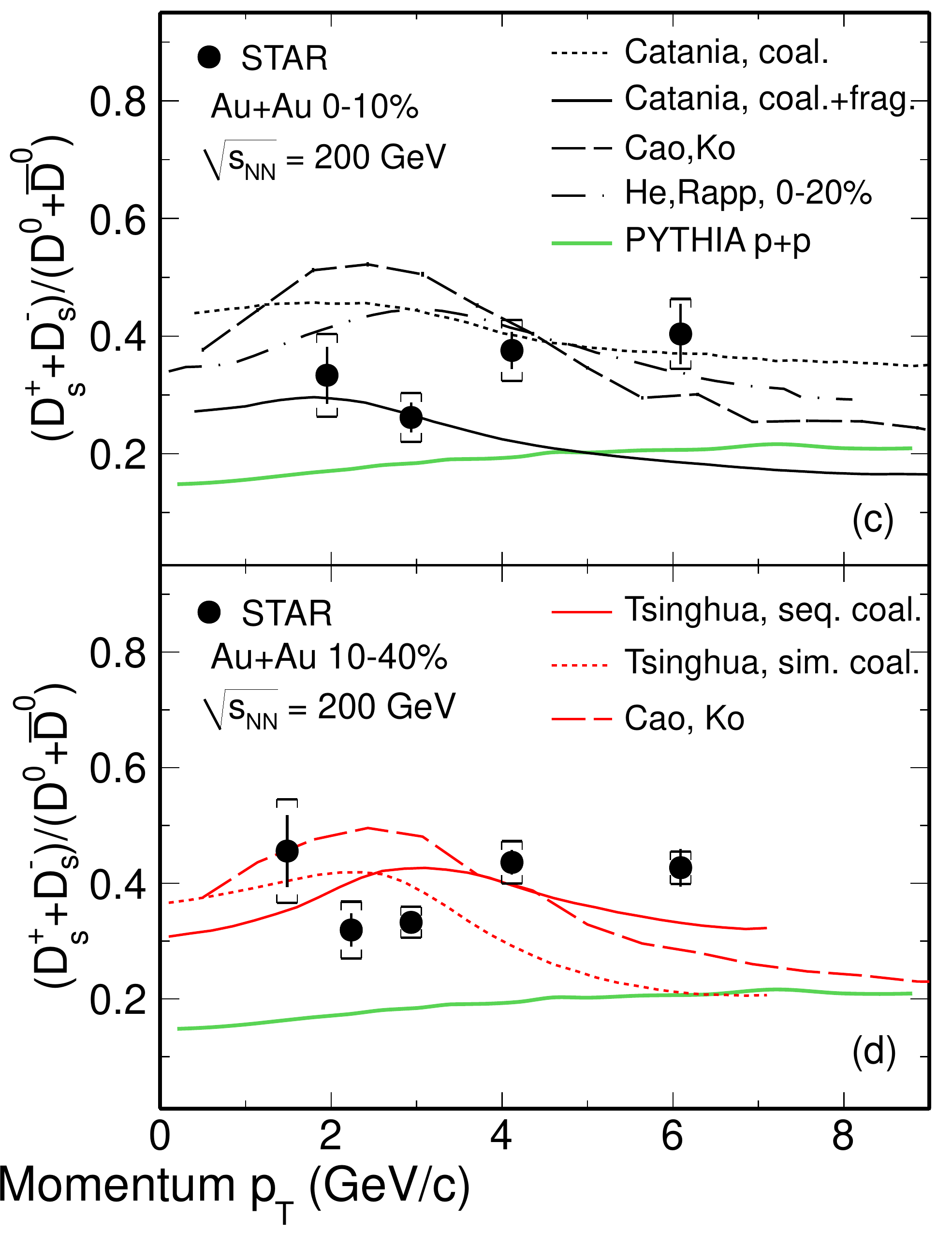}
\caption{(Color online) Left panels: The measured $\Lambda_{c}/D^0$ ratio at midrapidity ($|y|<$ 1) as a function of $p_{T}$ for Au+Au collisions at $\sqrt{s_{NN}}$ = 200\,GeV in 10\%--80\% centrality, compared to the baryon-to-meson ratios for light and strange hadrons (\textbf{a}) and various model calculations (\textbf{b}). The~$p_{T}$-integrated $\Lambda_{c}/D^0$ ratio from the THERMUS~\cite{Wheaton:2004qb} model calculation with a freeze-out temperature of $T_{\rm ch}=160$\,MeV is shown as a horizontal bar on the left axis of the plot. Right panels: (\textbf{c}) The integrated $D_{s}/D^{0}$ ratio (black solid circles) of 1.5 $<$ $p_{T}$ $<$ 8\,GeV/$c$ as a function of $p_T$ compared to model calculation (curves) in 0--10\%Au+Au collisions at $\sqrt{s_{NN}}$ = 200\,GeV. (\textbf{d}) Same $D_s/D^0$ ratio as (c) but with 10--40\% centrality. The vertical lines and brackets on the data points indicate statistical and systematic uncertainties respectively. }
\label{chadronization}
\end{figure}

The STAR Heavy Flavor Tracker (HFT) with a silicon pixel detector achieved $\sim$30 $\upmu$m spacial resolution of the track impact parameter to the primary vertex allows a topological reconstruction of the decay vertices of open charm hadrons. Figure~\ref{chadronization} left panels show the charmed baryon over meson ratio compared with light and strange baryon over meson ratios~\cite{Agakishiev:2011ar,Abelev:2006jr} (a) and various models~(b). The~$\Lambda_c/D^0$ ratio is comparable in magnitude to the $\Lambda/K^0_s$ and $p$/$\pi$ ratios and shows a similar $p_{T}$ dependence in the measured region. A~significant enhancement is seen compared to the calculations from the latest PYTHIA 8.24 release (Monash tune~\cite{Skands:2014pea}) without (green solid curve) and without (magenta dot-dashed curve) color reconnections (CR)~\cite{Bierlich:2015rha}. The~implementation with CR is found to enhance the baryon production with respect to mesons. However, both calculations fail to fully describe the data and its $p_{T}$ dependence. Figure~\ref{chadronization}b also shows the comparison to various models with coalescence hadronization of charm quarks~\cite{Oh2009,Plumari:2017ntm,Zhao:2018jlw,He2013}. The~comparisons suggest coalescence hadronization plays an important role in charm-quark hadronization in the presence of QGP. Also, the~data can be used to constrain the coalescence model calculations and their model~parameters. 

Figure~\ref{chadronization} right panel shows the $D_{s}/D^{0}$ ratio as a function of $p_{T}$ compared to coalescence model calculations for 0--10\% (c) and 10\%--40\% (d) collision centralities. Several models incorporating coalescence hadronization of charm quarks and strangeness enhancement are used to describe the $p_{T}$ dependence of $D_{s}/D^{0}$ ratio. 
Those models assume that $D_s^{\pm}$ mesons are formed by recombination of charm quarks with equilibrated strange quarks in the QGP~\cite{Oh2009,Plumari:2017ntm,Zhao:2018jlw,He2013}. In~particular, the~sequential coalescence model together with charm quark conservation~\cite{Zhao:2018jlw} considers that more charm quarks are hadronized to $D_s^{\pm}$ mesons than $D^{0}$ since the former is created earlier in the QGP, which results in further enhancement of $D_{s}/D^{0}$ ratio in Au+Au collisions relative to $p+p$ collisions.

$D$ meson $R_{AA}$ and $v_2$ have been observed similar as light flavor hadrons in 200 GeV Au+Au collisions~\cite{STARDAA}, which indicates charm is thermalized in the system with $T\sim$170 MeV. In~low energy region, such as the energies in RHIC beam energy scan program, it is of particular interest to measure open charm hadron production in a relative smaller and colder system compared to top energy at 200 GeV. This may provide a chance to tell us in what temperature charm behaves different from light flavors. However, in~low energy region, the~perturbation algorithm in theoretical calculations of charm production cross section becomes invalid, which may result in large theoretical uncertainties. Meanwhile the charm production cross section drops rapidly when collision energy decreases, it is very challenging to measure open charm production at low energies. The~previous measurements at SPS energies are with large uncertainties~\cite{Adc02,FMN97}. Since the HFT detector was taken out from STAR for the BES-II runs together with low cross section, it is impossible to reconstruct open charm hadrons via hadronic decay channels, the~electron production from heavy flavor semi-leptonic decays becomes the unique way to measure heavy flavor productions at low~energy. 

\begin{figure}[H]
\centering
\includegraphics[width=8 cm]{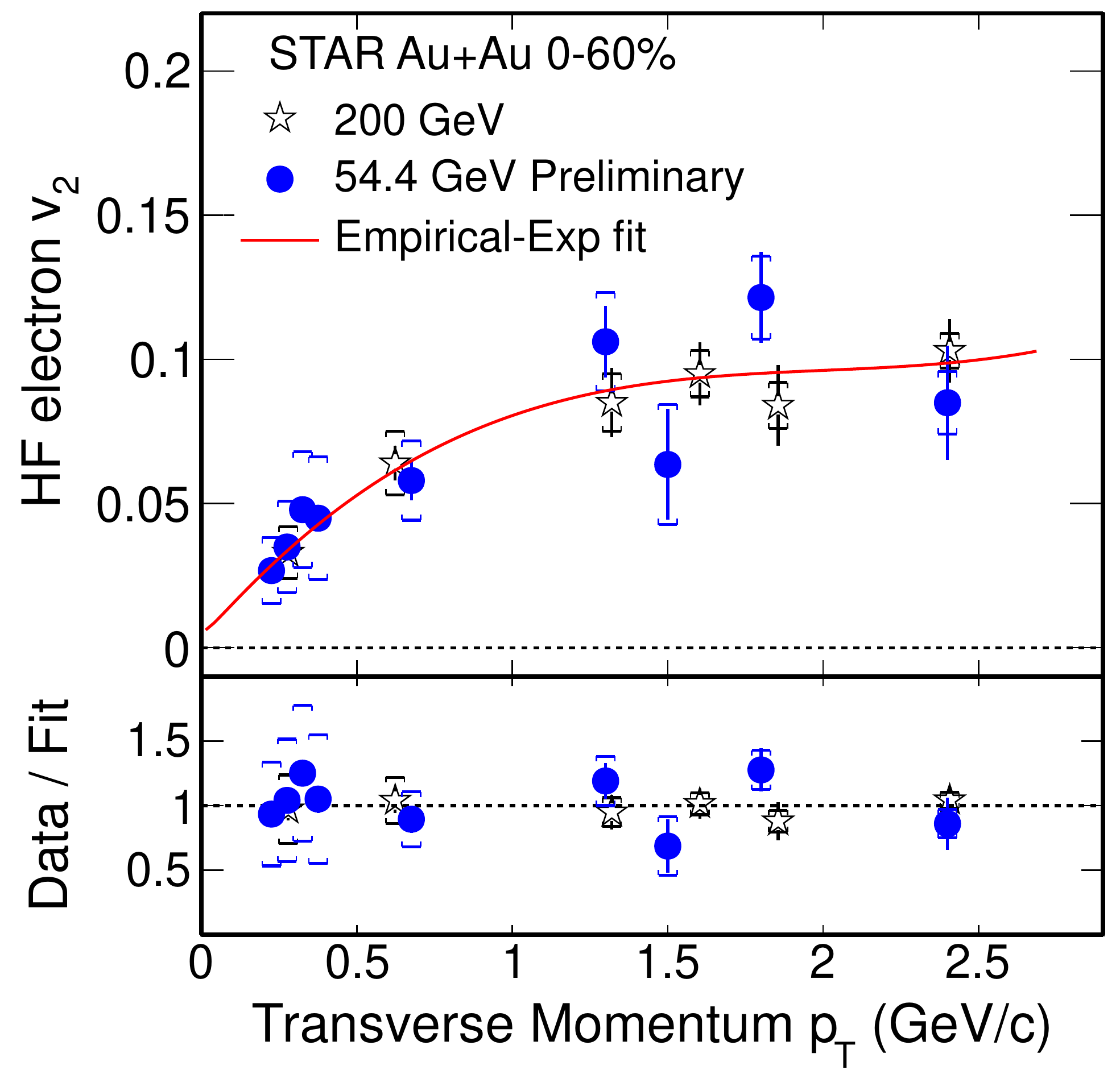}
\caption{(Color online) Upper panel: $v_2$ of electrons from heavy flavor decays as a function of $p_{T}$ in $\sqrt{s_{\rm NN}} =$ 54 (blue solid circles) and 200 (open stars) GeV Au+Au collisions~\cite{NPE62200}. Red curve denotes an empirical reversed exponential fit~\cite{ExpFuc} to all the data points. Bottom panel: The ratios of data over the fit function. Vertical bars and brackets denote statistical and systematic uncertainties, respectively.}
\label{npev254}
\end{figure}

Figure~\ref{npev254} shows the $v_2$ of electrons from heavy flavor decays as a function of $p_{T}$ in $\sqrt{s_{\rm NN}} =$ 54 and 200 GeV Au+Au collisions~\cite{NPE62200} as solid circles and open stars, respectively. A~semi-empirical exponential function~\cite{ExpFuc} is used to fit all the data points and the ratios of data over the fit function are shown in the bottom panel. The~result in 54 GeV agrees with that in 200 GeV within uncertainties, which may suggest charm quarks are still thermalized in 54 GeV Au+Au collisions. On~the other hand, it is of interest to repeat the same measurement in lower energies, such as 27 GeV from STAR BESII~experiment.

STAR experiment extracted the total charm production cross section per binary nucleon collision at midrapidity in $\sqrt{s_{\rm NN}} =$ 200 GeV Au+Au collisions by summing all yields of the open charm hadron states and reported as $d\sigma^{\rm NN}/dy|_{y=0}$ = 152 $\pm$ 13 (stat) $\pm$ 29 (sys) $\mu b$~\cite{STARccXesc}, which is consistent with that in $p$+$p$ collisions $d\sigma/dy|_{y=0}$ = 130 $\pm$ 30 (stat) $\pm$ 26 (sys) $\mu b$~\cite{STARDpp} within uncertainties. This result is consistent with charm quark conservation in heavy-ion collisions at RHIC top~energy.

\begin{figure}[H]
\centering
\includegraphics[width=8.1 cm]{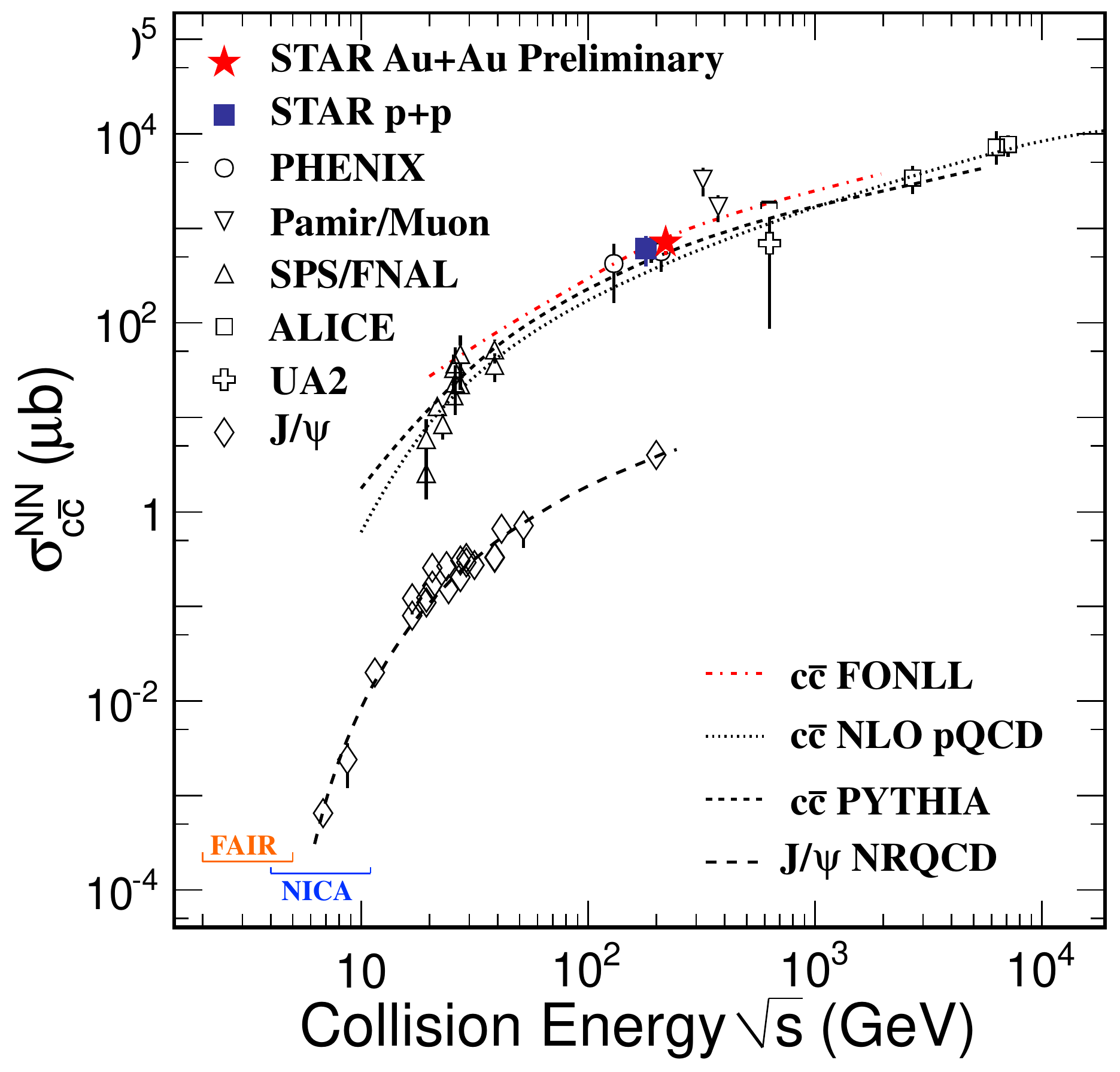}
\caption{(Color online) Charm and J/$\psi$ total production cross sections per nucleon-nucleon as a function of collision energy. The~open diamonds denote the charmonium cross section from worldwide experiments~\cite{rCERN-PS,rWA39,rIHEP,rE288,rE331,rE444,rE595,rE672,rE705,rE771,rE789,rNA3,rNA38,rNA50,rNA51,rUA6,rHERA-B,rISR,rPHENIX}. The~other open symbols are the experiment results of charm total cross section taken from Ref.~\cite{Tav87,Adc02,FMN97}. STAR $p$+$p$~\cite{STARDpp} and Au+Au~\cite{STARccXesc} results are shown as blue solid square and red star, respectively. Model calculations from FONLL~\cite{FONLL}, NLO pQCD~\cite{pQCD}, PYTHIA~\cite{pythia} and NRQCD~\cite{JpsiXesc} are represented as dot-dashed, dotted, dashed and long-dashed curves, respectively.}
\label{ctotxsec}
\end{figure}

The total charm production cross section in full rapidity region can be calculated from above charm cross section at midrapidity multiplying an equivalent correction factor (4.7 $\pm$ 0.7) assuming charm quark rapidity distribution from PYTHIA calculations~\cite{pythia}. Figure~\ref{ctotxsec} shows the charm total production cross section over a wide collision energy region from a few ten GeV to TeV. Open symbols are the experimental results taken from Ref.~\cite{Tav87,Adc02,FMN97}. STAR $p$+$p$~\cite{STARDpp} and Au+Au~\cite{STARccXesc} results are shown as blue solid square and red star, respectively. As~for comparison, the~total cross section of charmonium measured from CERN-PS~\cite{rCERN-PS}, WA39~\cite{rWA39}, IHEP~\cite{rIHEP}, E288~\cite{rE288}, E331~\cite{rE331}, E444~\cite{rE444}, E595~\cite{rE595}, E672~\cite{rE672}, E705~\cite{rE705}, E706~\cite{rE672}, E771~\cite{rE771}, E789~\cite{rE789}, NA3~\cite{rNA3}, NA38~\cite{rNA38}, NA50~\cite{rNA50}, NA51~\cite{rNA51}, UA6~\cite{rUA6}, HERA-B~\cite{rHERA-B}, ISR~\cite{rISR}, PHENIX~\cite{rPHENIX} experiments (open diamonds) and NRQCD (long-dashed curve) are shown as well over a broad collision energy region~\cite{JpsiXesc}.

\section{Future Upgrades and Physics Program at High Baryon Density~Region}
\begin{figure}[H]
\centering
\includegraphics[width=12 cm]{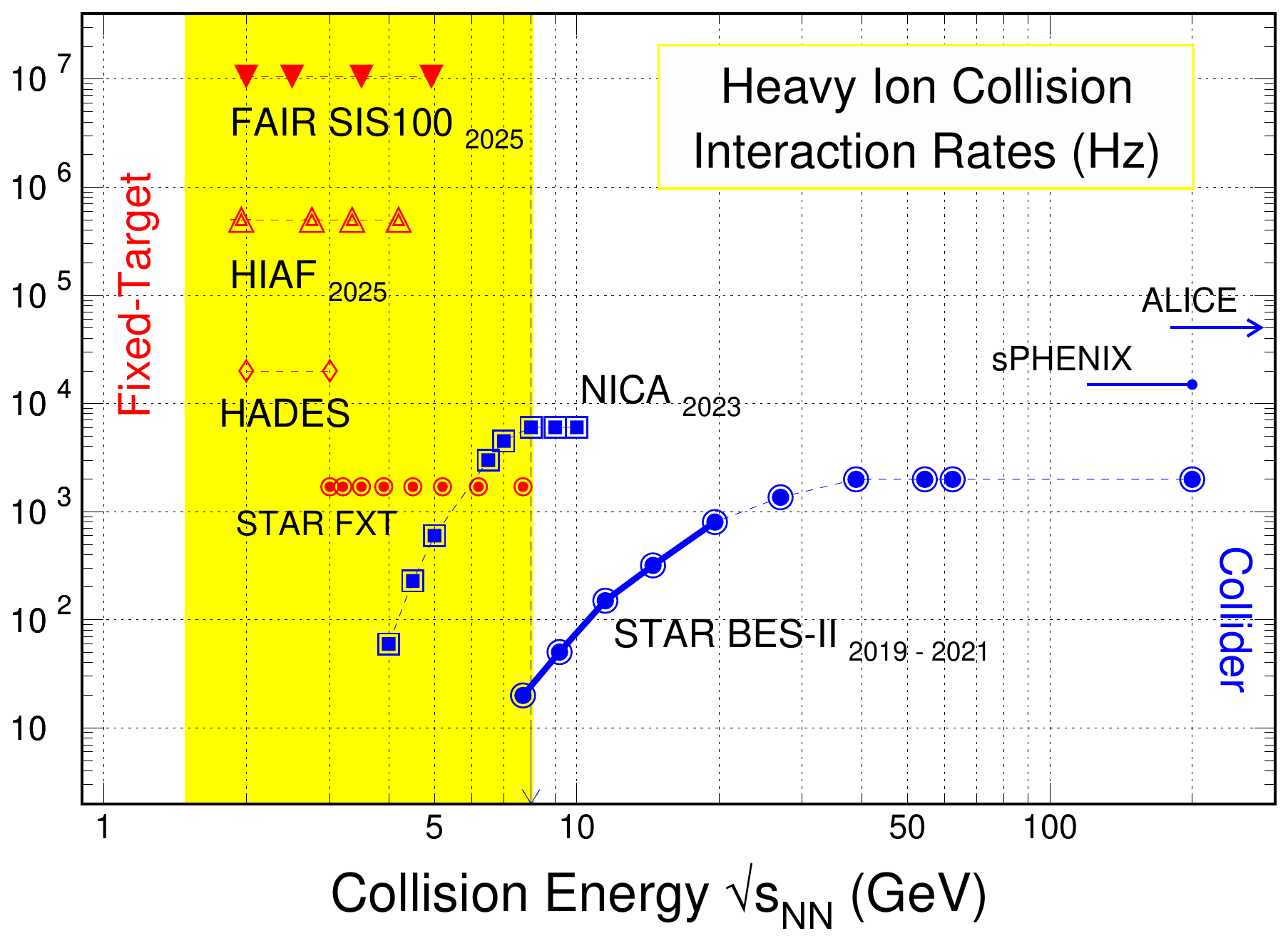}
\caption{(Color online) Interaction rates for high-energy nuclear collision facilities: the second phase RHIC beam energy scan (filled blue circles: BES-II, $7.7 < \sqrt{s_{NN}} < 19.6$ GeV), NICA (squares: $4 < \sqrt{s_{NN}} < 11$ GeV) as well as the fixed-target projects including HADES (diamonds), HIAF (triangles) and FAIR (filled triangles). 
STAR fixed-target range is indicated with filled red circles.}\label{fig16_ratesvsenergy_hi_23nov2019}
\end{figure}

The RHIC BES program II and future FAIR and NICA experiments will focus on collision energy below 20 GeV offering us a unique opportunity to explore 
the QCD phase structure at high baryon density region. In Figure~\ref{fig16_ratesvsenergy_hi_23nov2019}, interaction rates from both collider experiments and fixed-target experiments are shown. The~region for the high baryon density, largely covered by the fixed-target experiments, is highlighted with yellow. In~the following, we discuss few key measurements with the future experimental facilities. Again the discussions are arranged around the headlines of Collectivity, Criticality and Heavy Flavor~Productions. 

{\it Collectivity}: The flow results from top energy heavy-ion collisions at RHIC indicate that the partonic collectivity has been built up
from light $u$, $d$ and $s$ quarks to heavy $c$ quark as well. This is one of the most important experimental evidences for the creation of the QGP in high-energy nuclear collisions~\cite{BRAHMS:qgp, PHOBOS:qgp, STAR:qgp, PHENIX:qgp}. 
As a function of the collision energy, both radial and elliptic flow show an increasing trend, say above $\sqrt{s_{NN}} \approx 15$ GeV (Figure~\ref{v0v2}). Above~that energy, the~$v_1$ slope of net-particles for both baryons and mesons, is observed to be almost 
the same, the~$v_2$ difference between particle and anti-particle also becomes similar (Figure~\ref{BESflow}). While the $dv_1/dy$ shows large divergence between net-kaon and net-proton (and net-$\Lambda$), the~particle and anti-particle $v_2$ difference splits between baryons and mesons dramatically below
$\sqrt{s_{NN}} = 15$ GeV, see Figure~\ref{BESflow}. All of these observations imply that the medium properties created in heavy-ion collisions would be different above/below $\sqrt{s_{NN}} = 15$ GeV.

In collectivity, two noticeable observations are the splitting between baryon's and meson's $v_1$ and $v_2$ in the low energy. Mesons such as kaons and $\phi-$mesons are important, especially the $\phi-$meson as it has the similar mass of proton. The~precise results of $\phi-$meson's $v_1$ and $v_2$ will reveal the origin of collectivity at the high baryon density region. In~addition, the~ratio of $N(K^-)/N(\phi)$ will shed light on the production mechanism. It could be treated as a micro-laboratory for understanding the quarkonia productions in nucleus-nucleus~collisions. 

{\it Criticality}: One of the main goal of RHIC Beam Energy Scan program is to search for the QCD critical point, which is the end point of the first order phase boundary in the QCD phase diagram. The~experimental confirmation of the existence of the CP will be a landmark of exploring the QCD phase structure. Near~the QCD critical point, the~density fluctuations and correlation length will diverge. The~conserved charge fluctuations and light nuclei productions have been proposed as sensitive observables to search for the signature of QCD phase transition and the QCD critical point. \textls[-5]{Experimentally, the~STAR experiment has measured the higher order cumulants of net-particle distributions and light nuclei productions (deuteron and triton) in Au+Au collisions at \mbox{$\sqrt{s_{NN}}$ = 7.7}} to 200~GeV. In~Figures~\ref{fig:KV_CBM} and~\ref{fig:light_nuclei}, it has been observed that both fourth order fluctuations of net-proton ($\kappa \sigma^{2}$) and light nuclei yield ratio $N_t \times N_p /N^{2}_d$ show non-monotonic energy dependence in central Au+Au collisions, with~a minimum and peak around 20 GeV, respectively. Although~the two measurements are of different order of fluctuations, the~two observations are consistent with the expectation of model calculations with CP physics and might suggest that the created system skims close by the CP receiving the contributions from critical fluctuations. To~confirm the above two observed non-monotonic energy dependence trends in BES-I, the~second phase of Beam Energy Scan (BES-II) has been planned at RHIC (2019--2021). It will allow us to have 10--20 times more statistics at energies $\sqrt{s_{NN}}$ = 7.7--19.6 GeV.  
In~addition, one observes large changes between 19.6 and 14.5 GeV in the energy dependence of net-proton kurtosis (Figure~\ref{fig:KV_CBM}) and light nuclei yield ratio $N_t \times N_p /N^{2}_d$ (Figure~\ref{fig:light_nuclei}) measured in the RHIC BES-I data. This could indicate that the QCD critical point is put by nature between the thermodynamic condition ($T, \mu_B$) of 19.6 and 14.5 GeV. Thus, it is important to conduct a finer beam energy scan between these two energies, i.e.,~19.6 GeV ($\mu_B=205$ MeV) and 14.5 GeV ($\mu_B=266$ MeV). Therefore, we propose to take the data of a new energy point of Au+Au collisions at $\sqrt{s_{NN}}$ = 16.7 GeV ($\mu_B=235$ MeV), which is just between 19.6 and 14.5 GeV with equal $\mu_B$ gap, on~each side. Based on the net-proton fluctuations measured from HADES and STAR experiments, and~the model calculations, there might be a peak in the fourth order net-proton fluctuations in Au+Au collisions between \sNN\ = 2 GeV and 8 GeV. In~order to experimentally map out the QCD phase diagram at the higher baryon density region, the~future heavy-ion collision experiments like MPD/NICA, CBM/FAIR and CEE/CSR are certainly necessary and~important. 

On the other hand, it is predicted that the higher order conserved charge fluctuations, such as sixth order ($C_6$) or eighth order ($C_8$) cumulants, should be more sensitive to the phase transition. If~the chemical freeze-out temperature in heavy-ion collisions are close enough to the phase boundary, the~sixth and eighth order fluctuations could show negative values~\cite{Friman:2011pf,Bazavov:2020bjn}. STAR experiment has measured the centrality dependence of sixth order ($C_6/C_2$) of net-proton distributions in Au+Au collisions at $\sqrt{s_{NN}}$ = 54.4 and 200 GeV. Negative values are observed for net-proton $C_6/C_2$ from mid-central to central collisions at 200 GeV, while positive values are observed at 54.4 GeV~\cite{Nonaka_QM2019,Ashish_QM2019}. The~negative sign of net-proton $C_6/C_2$ observed at 200 GeV could be an experimental evidence of smooth crossover at small baryon chemical potential~\cite{Friman:2011pf,Bazavov:2020bjn}. In~future fixed target experiments, with~much more statistics of low energy data, we can perform precise measurements of those higher order cumulants of conserved charges at the high baryon density~region.

{\it Heavy Flavor Production}: FAIR-CBM and NICA-MPD experiments with advanced fast detector technology under high luminosity beam condition will 
provide unique chance to measure open and hidden charm hadrons with large statistics close to the production energy threshold~\cite{Ablyazimov:2017guv,MPD19}. 
It is expected that this measurements will improve the precision of the total charm cross section at low energies and will provide constraints on 
pQCD calculations, as~well as the unknown interactions between charmed particles and cold hadronic medium. 
Taking the prediction of the HSD model~\cite{HSD01}, the~yield obtained in one week of running of CBM detectors with 
10 MHz event rate would be about 300 J/$\psi$ for central Au+Au collisions at 10A GeV, and~about 600 J/$\psi$ 
for central Ni+Ni collisions at 15A GeV. In~the latter case, also open charm production can be studied at a rate of 300 kHz 
with a silicon vertex detector MVD in operation for charmed hadron decay vertex reconstruction. 
As a result, the~expected yield in central Ni+Ni collisions at 15A GeV will be about 30 $D$ mesons per week. 
This would be sufficient for cross section measurement and an analysis of charmonium propagation and absorption 
in dense baryonic matter based on the ratio of hidden to open charm at low~energy.

In order to extend the coverage to even larger baryon density region, STAR has developed a fixed-target (FXT) program. 
As shown in Figure~\ref{fig16_ratesvsenergy_hi_23nov2019}, a~gold-target (1\% interaction length) is placed at the right entrance of TPC. 
The end cap time-of-flight wall will be constructed at approximately the other side of the TPC entrance. 
\textls[-15]{The time-of-flight detectors are on loan from the CBM experiment at FAIR~\cite{future1, future2}. 
In addition, the~inner-TPC upgrade~\cite{future3} will extend the rapidity coverage, essential for the search for the QCD critical point measurement. 
STAR will be setup in such a way that data taking from both colliding and FXT modes will take place concurrently. 
With this configuration, STAR detector system will measure particle productions and correlations in Au+Au collisions from \mbox{$\sqrt{s_{NN}}$ = 3--19.6 GeV}}, 
extending its coverage of baryon chemical potential from about $\mu_B$ = 400 MeV to $\mu_B \sim$ 750 MeV. 
The center of mass energy from the highest energy of the FXT mode is overlay with the lowest colliding mode at $\sqrt{s_{NN}}$ = 7.7 GeV 
and the lower part of the FXT energies overlap with the future collision energies provided by CBM at FAIR~\cite{cbm}. 
These allow systematic crosschecks on many of the observables in STAR experiment, for~both colliding and FXT modes, 
and CBM experiment for the FXT mode. The~BES program II and future FAIR and NICA experiments will focus on the high baryon density region (<20 GeV), 
offer us a unique opportunity to explore the QCD phase~structure. 

In summary, the precise flow measurements of $\phi$ mesons and multi-strange hadrons with STAR BES-II and future fixed-target experiments will reveal the 
degree of freedom originates from partonic or hadronic level at the high baryon density region. The~confirmation of non-monotonic energy dependence in central Au+Au collisions for net-proton $\kappa \sigma^{2}$ and/or light nuclei yield ratio $N_t \times N_p /N^{2}_d$ with STAR BES-II and future fixed-target experiments will provide crucial experimental evidences for establishing the case for the discovery of the QCD critical point. The~energy dependence of heavy flavor measurements will provide crucial information on thermalization of the system and provide unique opportunity to study the unknown interactions between heavy quark and the cold nuclear matter. A~great deal of new information on the QCD phase diagram will be extracted with current and planned heavy-ion collision~programs.


\vspace{6pt} 



\authorcontributions{X.L., S.S., N.X. and Y.Z. contribute equally to this paper. All authors have read and agreed to the published version of the manuscript.}
\funding{This work is supported by the National Key Research and Development Program of China (2018YFE0205200),  the~National Natural Science Foundation of China (No.11890711, 11890712, 11828501 and 11861131009).}

\acknowledgments{We thank Xin Dong, ShinIchi Esumi, Lokesh Kumar, Volker Koch, Bedangadas Mohanty for~discussions.}

\conflictsofinterest{The authors declare no conflicts of interest.}

\reftitle{References}






\end{document}